\documentclass[prd,preprint,amssymb,nofootinbib,superscriptaddress,letterpaper]{revtex4}
\usepackage[dvips]{graphicx}
\usepackage{setspace}
\begin{document}
\preprint{hep-ph/0612317}

\title{Physics impact of ILC Higgs coupling measurements: \\
the effect of theory uncertainties}

\author{Andrew Droll}

\author{Heather E.\ Logan}
\email{logan@physics.carleton.ca}

\affiliation{Ottawa-Carleton Institute for Physics,
Carleton University, Ottawa K1S 5B6 Canada}

\begin{abstract}
We study the effect of theoretical and parametric uncertainties on the 
ability of future Higgs coupling measurements at the International Linear
Collider (ILC) to reveal deviations from the Standard Model (SM).  To quantify
the impact of these uncertainties we plot 
$\Delta \chi^2 = 25$ contours for the deviations between the SM Higgs 
couplings and the light Higgs couplings in the $m_h^{\rm max}$ benchmark 
scenario of the Minimal Supersymmetric Standard Model (MSSM).
We consider the theoretical uncertainties in the SM Higgs decay partial widths
and production cross section and the parametric uncertainties in the bottom
and charm masses and the strong coupling $\alpha_s$.  
We find that the impact of the theoretical and parametric uncertainties is 
moderate in the first phase of ILC data-taking (500 fb$^{-1}$ at 350 GeV 
centre-of-mass energy), reducing the ``reach'' in the CP-odd MSSM Higgs
mass $M_A$ by about 10\% to $\sim 500$~GeV, 
while in the second phase (1000 fb$^{-1}$ at 1000 
GeV) these uncertainties are larger than the experimental uncertainties and
reduce the reach in $M_A$ by about a factor of two, from $\sim 1200$ down
to $\sim 600$~GeV.  The bulk of the effect comes from the parametric 
uncertainties in $m_b$ and $\alpha_s$, followed by the theoretical 
uncertainty in $\Gamma_b$.
\end{abstract}

\maketitle

\section{Introduction}

An important component of the physics case for the International Linear
Collider (ILC) is its ability to perform high-precision measurements in the
Higgs sector~\cite{TeslaTDR,OrangeBook,Snowmass05Higgs}.
In addition to measuring the Higgs mass, its spin and CP quantum numbers, 
and its self-coupling which characterizes the shape of the Higgs potential,
the ILC will measure the Higgs couplings to a variety of Standard Model (SM) 
particles.  In the SM, the masses of these particles arise solely through 
their couplings to the Higgs, so that these couplings are predicted
in terms of the corresponding SM particle masses.  
The SM Higgs production cross
sections and decay branching ratios are then fixed once the Higgs mass has 
been measured.

In extended models, however, the masses of the SM particles can receive 
contributions from more than one source, e.g., from couplings to each of two 
Higgs doublets.  Further, the SM-like Higgs boson in extended models is 
typically an admixture of states from different scalar multiplets, leading
to modifications of its couplings to SM particles by additional mixing angles.
The high-precision ILC measurements of Higgs couplings thus provide a 
powerful tool for probing the structure of the Higgs sector and distinguishing
the minimal SM from extended models.

Previous
studies~\cite{BattagliaMA,CHLM,OkadaMA,PenarandaDb,DawsonHeinemeyer}
have quantified the ``reach'' of the ILC measurements within the
context of the Minimal Supersymmetric Standard Model (MSSM).  In this
paper we study the impact on this reach due to the theoretical and
parametric uncertainties that enter the SM predictions for the Higgs
cross section and decay branching ratios.  A first attempt at
including the parametric uncertainties from the strong coupling and
the bottom and charm quark masses was made in Ref.~\cite{CHLM} by
varying these parameters within their uncertainties and treating the
resulting variation in the SM Higgs branching ratio predictions as an
additional uncertainty to be added in quadrature with the uncertainty
on the experimental measurement.  An additional step was taken in
Ref.~\cite{LHatPC} where the theoretical uncertainties in the SM
$\gamma\gamma \to H \to b \bar b$ rate were estimated based on the
size of the computed radiative corrections.  We refine and extend the
analysis of the parametric uncertainties by taking into account their
correlated effects on Higgs observables.  We also include the full set
of current theoretical uncertainties in the SM Higgs decay partial
widths and production cross section, again taking into account their
correlated effects on the ILC Higgs observables.

With the theoretical and parametric uncertainties in hand, we must choose 
a measure to quantify their impact.  We choose to work in the $m_h^{\rm max}$
benchmark scenario of the MSSM as defined in Ref.~\cite{CarenaBenchmarks}.
We scan over points in the $M_A$--$\tan\beta$ plane and construct a 
chi-squared ($\chi^2$) observable between the SM and MSSM predictions.  We plot
contours of $\Delta \chi^2 = 25$, corresponding to a $5 \sigma$ discrepancy 
between the SM and the sample MSSM point.  These contours allow us to show 
the impact of adding each of the theoretical and parametric uncertainties on
top of the experimental uncertainties.
Because the main purpose of this paper is to examine the impact of the
theoretical uncertainties, we make no attempt to scan over more general sets 
of MSSM parameters or over non-supersymmetric models.

We begin in the next section with a brief overview of the MSSM Higgs
couplings.  framework of our analysis.  In Sec.~\ref{sec:chisquared}
we describe our $\chi^2$ procedure for dealing with the uncorrelated
and correlated sources of uncertainty.  In Sec.~\ref{sec:inputs} we
collect the expected experimental uncertainties in the Higgs
measurements and describe our treatment of the theoretical and
parametric uncertainties, which typically feed in to multiple
measurements in a correlated way.  We provide a table of the
dependence of individual Higgs decay partial widths on the input
parameters $m_b$, $m_c$ and $\alpha_s$ so that our study can be
updated in a straightforward way if the precision on these parameters
improves.  We present our numerical results in Sec.~\ref{sec:results}.
Section~\ref{sec:conclusions} is reserved for discussion and
conclusions.  The current state of the art for the radiative
corrections to SM Higgs decays, Higgs production in $e^+e^- \to \nu
\bar \nu H$, and the bottom and charm quark mass extraction from
low-energy experimental data is reviewed in Appendices~\ref{app:thy},
\ref{app:xsec}, and \ref{app:mbmc}, respectively.

\section{Theoretical framework}
\label{sec:theory}

The couplings of the lightest CP-even Higgs boson $h^0$ of the MSSM 
can be written at tree level in terms of the corresponding SM Higgs 
couplings as 
\begin{eqnarray}
  \frac{g_{h^0 \bar t t}}{g_{H_{\rm SM} \bar t t}} = 
  \frac{g_{h^0 \bar c c}}{g_{H_{\rm SM} \bar c c}} &=& 
  \sin(\beta - \alpha) + \cot\beta \cos(\beta - \alpha) 
  \nonumber \\
  \frac{g_{h^0 \bar b b}}{g_{H_{\rm SM} \bar b b}} =
  \frac{g_{h^0 \tau \tau}}{g_{H_{\rm SM} \tau \tau}} &=& 
  \sin(\beta - \alpha) - \tan\beta \cos(\beta - \alpha) 
  \nonumber \\
  \frac{g_{h^0 WW}}{g_{H_{\rm SM} WW}} =
  \frac{g_{h^0 ZZ}}{g_{H_{\rm SM} ZZ}} &=& \sin(\beta - \alpha),
  \label{eq:higgscoups}
\end{eqnarray}
where $\alpha$ is the mixing angle that diagonalizes the mass matrix for the
CP-even states $h^0$ and $H^0$, and
$\tan\beta = v_2/v_1$ is the ratio of vacuum expectation values of the 
two MSSM Higgs doublets.  At tree level we have
\begin{equation}
  \cos(\beta - \alpha) \simeq \frac{1}{2} \sin 4\beta \frac{m_Z^2}{M_A^2},
  \label{eq:decoupling}
\end{equation}
where $M_A$ is the mass of the CP-odd neutral MSSM Higgs boson.
Note that in the \emph{decoupling limit}
$M_A \gg m_Z$~\cite{HowieDecoup}, $\cos(\beta - \alpha)$
goes to zero and the $h^0$ couplings approach their SM values.
Using 
Eqs.~(\ref{eq:higgscoups}) and (\ref{eq:decoupling}) to expand the Higgs 
partial widths in powers of $m_Z^2/M_A^2$, we obtain~\cite{CHLM}
\begin{eqnarray}
  \frac{\delta \Gamma_W}{\Gamma_W} = \frac{\delta \Gamma_Z}{\Gamma_Z} 
  &\simeq& - \frac{1}{4} \sin^2 4\beta \frac{m_Z^4}{M_A^4}
  \simeq - 4 \cot^2 \beta \frac{m_Z^4}{M_A^4}  
  \nonumber \\
  \frac{\delta \Gamma_b}{\Gamma_b} 
  \simeq \frac{\delta \Gamma_{\tau}}{\Gamma_{\tau}} 
  &\simeq& - \tan\beta \sin 4\beta \frac{m_Z^2}{M_A^2}
  \simeq + 4 \frac{m_Z^2}{M_A^2} 
  \nonumber \\
  \frac{\delta \Gamma_c}{\Gamma_c}
  &\simeq& \cot \beta \sin 4\beta \frac{m_Z^2}{M_A^2}
  \simeq - 4 \cot^2 \beta \frac{m_Z^2}{M_A^2},
  \label{eq:deviations}
\end{eqnarray}
where the last equality in each line uses the large $\tan\beta$
approximation $\sin 4 \beta \simeq - 4 \cot\beta$.  In particular, 
for large $\tan\beta$, $\Gamma_b$ and $\Gamma_{\tau}$ exhibit the largest
deviations from their SM values, while $\Gamma_W$ approaches its SM value
very quickly with increasing $M_A$.

Beyond tree level, the Higgs couplings receive radiative corrections,
which mainly impact the CP-even Higgs mass
matrix~\cite{CarenaHaberReview}.  These can be taken into account by
defining an effective mixing angle $\alpha_{\rm eff}$ for use in place
of $\alpha$ in the above formulas.  Significant radiative corrections
can also appear in the $h^0 \bar b b$ vertex due to squark-gluino
loops, with their dominant piece parameterized as
$\Delta_b$~\cite{CarenaHaberReview}.  Including this vertex
correction, the $h^0 \bar b b$ coupling can be written as~\cite{CHLM}
\begin{equation}
  \frac{g_{h^0 \bar b b}}{g_{H_{\rm SM} \bar b b}} = 
  \left[ \sin(\beta - \alpha) - \tan\beta \cos(\beta - \alpha) \right]
  \frac{1 - \Delta_b \cot\alpha \cot \beta}{1 + \Delta_b}.
\end{equation}
In the limit of large $M_A$, $\cot\alpha \cot\beta \to -1$ and the 
$\Delta_b$ corrections also decouple.

In this paper we make our $\chi^2$ comparisons using the
implementations of the SM and MSSM Higgs couplings in the public
Fortran code {\tt HDECAY}~\cite{HDECAY}.  {\tt HDECAY} computes the
$h^0$ couplings using the $\alpha_{\rm eff}$ approximation,
incorporating also the potentially significant $\Delta_b$ corrections
to the $h^0 b \bar b$ coupling.  For the loop-induced couplings of
$h^0$ to photon or gluon pairs, {\tt HDECAY} includes the shift in the
contribution from SM particles in the loop\footnote{We added the charm
quark loop to the SM $\Gamma_{gg}$ calculation in {\tt HDECAY}, to be
consistent with its inclusion in the MSSM calculation.}  due to the
modified couplings in Eq.~(\ref{eq:higgscoups}), as well as the new
contributions to the amplitude from supersymmetric particles running
in the loop and from the charged Higgs boson for $h^0 \to
\gamma\gamma$.
For comparisons involving MSSM Higgs production cross sections in $WW$ 
fusion and $ZH$ associated
production, we scale the corresponding SM production cross section by 
$g^2_{h^0VV}/g^2_{H_{\rm SM}VV}$ ($V = W$ or $Z$) using the $\alpha_{\rm eff}$ 
approximation of {\tt HDECAY}.  We neglect the additional non-universal 
SUSY electroweak radiative corrections (e.g., box diagrams) beyond this
approximation.  
While a full MSSM analysis should include complete non-universal SUSY
corrections to Higgs partial widths and production cross sections,
the approximations used in {\tt HDECAY} are sufficient
for our purposes in this paper.

\section{Chi-squared analysis}
\label{sec:chisquared}

To quantify the impact of the experimental and theoretical uncertainties
on the ability of ILC measurements to reveal deviations from the SM, we 
construct a $\chi^2$ according to
\begin{equation}
  \chi^2 = \sum_{i=1}^{n}\sum_{j=1}^n (Q_i^{M_1} - Q_i^{M_2}) 
  [\sigma^2]^{-1}_{ij} (Q_j^{M_1} - Q_j^{M_2}),
\end{equation}
where $Q_i$ are the observables being compared between models $M_1$ and $M_2$,
and $[\sigma^2]^{-1}_{ij}$ is the inverse of the covariance matrix for the
quantities $Q_i$, defined according to
\begin{equation}
  \sigma^2_{ij} = \delta_{ij} u_i u_j + \sum_{k=1}^m c_i^k c_j^k.
\end{equation}
Here $u_i$ is the uncorrelated uncertainty in the observable
$Q_i$, $c_i^k$ represents the $k$th source of correlated uncertainty in 
$Q_i$, and $\delta_{ij}$ is the Kronecker delta.  In the absence
of correlated uncertainties, $\sigma^2_{ij}$ is a diagonal matrix with the 
(positive definite) squares of the uncorrelated uncertainties $u_i$ down the 
diagonal.  Correlated uncertainties introduce off-diagonal terms, which can
be positive or negative depending on the sense of the correlation 
between observables $Q_i$ and $Q_j$.

In our analysis the observables $Q_i$ consist of Higgs branching fractions
and Higgs production cross sections times branching fractions (i.e., rates
in a particular channel).  We take the expected experimental uncertainties
on these quantities at the ILC from the literature.  Because the existing
experimental studies have not quantified the correlations between the measured
observables, we treat the experimental uncertainties as uncorrelated.  
The theoretical and parametric uncertainties affect multiple observables,
and we therefore treat them as correlated uncertainties.

\section{Inputs}
\label{sec:inputs}

\subsection{Experimental inputs}

We consider Higgs measurements from two stages of ILC running:
\begin{itemize}

\item {\bf Phase 1:} 500 fb$^{-1}$ of integrated luminosity at 350 GeV 
centre-of-mass energy, with no beam polarization.
Expected uncertainties in the SM Higgs branching ratios for Higgs masses
$m_H = 120$ and 140 GeV are taken from Ref.~\cite{Desch} and summarized 
in Table~\ref{tab:p1uncert}.

\item {\bf Phase 2:} 1000 fb$^{-1}$ of integrated luminosity at 1000 GeV
centre-of-mass energy, with beam polarizations of $-80\%$ for 
electrons and $+50\%$ for positrons.
Expected uncertainties in the SM Higgs production cross section times 
branching ratios for $m_H = 115$, 120 and 140 GeV are taken
from Ref.~\cite{Barklow} and summarized in Table~\ref{tab:p2uncert}.
For the Phase 2 analysis, we include the existing Phase 1 measurements in 
the $\chi^2$ in addition to the new measurements at 1000 GeV.  

\end{itemize}

\begin{table}
\begin{tabular}{ccc}
\hline \hline
\multicolumn{3}{c}{SM Higgs branching ratio uncertainties 
from 500 fb$^{-1}$ at 350 GeV} \\
 & $m_H = 120$ GeV & 140 GeV \\
\hline
BR($b \bar b$) & $2.4\%$ & $2.6\%$ \\
BR($c \bar c$) & $8.3\%$ & $19.0\%$ \\
BR($\tau\tau$) & $5.0\%$ & $8.0\%$ \\
BR($WW$) & $5.1\%$ & $2.5\%$ \\
BR($gg$) & $5.5\%$ & $14.0\%$ \\
\hline\hline
\end{tabular}
\caption{Expected fractional experimental uncertainties in the SM Higgs 
branching ratios for $m_H = 120$ and 140 GeV, from Ref.~\cite{Desch}
(500 fb$^{-1}$ at 350 GeV centre-of-mass energy with 
no beam polarization).}
\label{tab:p1uncert}
\end{table}

\begin{table}
\begin{tabular}{cccc}
\hline \hline
\multicolumn{4}{c}{SM Higgs cross section times BR statistical uncertainties 
from 1000 fb$^{-1}$ at 1000 GeV} \\
 & $m_H = 115$ GeV& 120 GeV & 140 GeV \\
\hline
$\sigma \times {\rm BR}(b \bar b)$ & $0.3\%$ & $0.4\%$ & $0.5\% $ \\
$\sigma \times {\rm BR}(WW)$ & $ 2.1\%$ & $1.3\%$ & $ 0.5\% $ \\
$\sigma \times {\rm BR}(gg)$ & $ 1.4\%$ & $1.5\%$ & $ 2.5\% $ \\
$\sigma \times {\rm BR}(\gamma\gamma)$ & $ 5.3\%$ & $5.1\%$ & $ 5.9\% $ \\
\hline\hline
\end{tabular}
\caption{Expected fractional experimental uncertainties in the SM Higgs
production cross section times decay branching ratios
for $m_H = 115$, 120 and 140~GeV, from Ref.~\cite{Barklow}
(1000 fb$^{-1}$ at 1000 GeV centre-of-mass energy).  Uncertainties are 
statistical only.  Beam polarizations of $-80\%$ for electrons and $+50\%$ 
for positrons are assumed.}
\label{tab:p2uncert}
\end{table}

The centre-of-mass energy of 350 GeV was chosen for the study in 
Ref.~\cite{Desch} because it is near the peak of the $e^+e^- \to ZH$ cross 
section for the Higgs masses considered.  It is also near the top quark pair 
production threshold, so that the Higgs data could be collected simultaneously
with a top threshold scan.  Running with polarized beams would improve the 
results by boosting the cross section and/or suppressing backgrounds.
The actual ILC run plan will of course depend on what is discovered at the 
CERN Large Hadron Collider (LHC), 
and may include multiple threshold scans or running at the maximal 
first-phase design energy of 500 GeV.

The study in Ref.~\cite{Barklow} at 1000~GeV centre-of-mass energy 
found the expected statistical uncertainties
on Higgs rates in various channels.  The analysis selected events
with $e^+e^- \to \nu \bar \nu H$, thus including Higgs production 
via $WW$ fusion and via $e^+e^- \to ZH$ with $Z \to \nu \bar \nu$.  The
Higgs decay products were then selected with the requirement that the
visible energy in the event add up to the Higgs mass.
The study in Ref.~\cite{Barklow} evaluated the statistical uncertainties
only.  In our $\chi^2$ analysis we add an overall luminosity uncertainty 
of 0.1\%~\cite{TeslaTDR} for the Higgs rate measurements at 1000~GeV, 
completely correlated among the four channels.
In this high-energy phase of ILC running, Higgs production is dominated by 
$WW$ fusion, the cross section for which grows with centre-of-mass energy for
a fixed $m_H$.  It is thus advantageous to run at the highest possible 
collider energy to maximize the statistics.  Beam polarization was chosen
to maximize the $WW$ fusion cross section.

In our analysis we scan over the $M_A$-$\tan\beta$ plane in the $m_h^{\rm max}$
scenario of the MSSM.  At each parameter point we take the Higgs observables 
for $h^0$ and those for a SM Higgs boson of the same mass as the input for our
$\chi^2$.  For Higgs masses between 120 and 140~GeV, we obtain the expected
experimental uncertainty on the Higgs observables using linear interpolation
between the values given in Tables~\ref{tab:p1uncert} and \ref{tab:p2uncert}.
In the $m_h^{\rm max}$ scenario the $h^0$ mass never exceeds 130~GeV.  For
$h^0$ masses below 120~GeV, which occur in this scenario only at low 
$\tan\beta$ values $< 4$, 
we use the experimental uncertainties for $m_H = 120$~GeV from
Table~\ref{tab:p1uncert} and apply linear interpolation only to the 
values given in Table~\ref{tab:p2uncert} between $m_H = 115$ and 120~GeV.

\subsection{Theoretical and parametric uncertainties}

\subsubsection{Higgs decay partial widths}

The theoretical uncertainties in the SM Higgs decay partial widths due
to uncalculated higher order radiative corrections are summarized in
Table~\ref{tab:thyuncert}.  We estimated these numbers based on the
size of the known higher order corrections for Higgs masses in our range
of interest; for details see Appendix~\ref{app:thy}.
In all cases we use the best theory uncertainty available in the 
literature for our $\chi^2$ calculation; 
in some cases the theory uncertainty in the {\tt HDECAY} 
calculation is larger because not all available radiative corrections have
been implemented into the {\tt HDECAY} code~\cite{Spirareview}.

\begin{table}
\begin{tabular}{ccc}
\hline \hline
  & \multicolumn{2}{c}{Theory uncertainty} \\
Higgs partial width & in literature & \hspace*{0.2cm} in {\tt HDECAY} \\
\hline
$\Gamma_{b \bar b}$, $\Gamma_{c \bar c}$ & 1\% & 1\% \\
$\Gamma_{\tau\tau}$, $\Gamma_{\mu\mu}$ & 0.01\% & 0.01\% \\
$\Gamma_{WW}$, $\Gamma_{ZZ}$ & 0.5\% & 5\% \\
$\Gamma_{gg}$ & 3\% & 16\% \\
$\Gamma_{\gamma\gamma}$ & 0.1\% & 4\% \\
$\Gamma_{Z\gamma}$ & 4\% & 4\% \\
\hline
Higgs production cross section & & \\
$\sigma_{e^+e^- \to \nu \bar \nu H}$ & 0.5\% & -- \\
\hline\hline
\end{tabular}
\caption{Estimated fractional theoretical uncertainties in the SM Higgs 
partial widths and production cross section due to uncalculated higher order 
corrections.
See Appendices~\ref{app:thy} and \ref{app:xsec} for details.}
\label{tab:thyuncert}
\end{table}

Working entirely in terms of fractional uncertainties,
the correlated uncertainty $c_i^k$ in branching ratio $i$ due to the 
theoretical uncertainty in the SM prediction for partial width $k$ is 
given by the formula
\begin{equation}
  c_i^k = \frac{\Gamma_k}{{\rm BR}_i} 
  \frac{\partial {\rm BR}_i}{\partial \Gamma_k} \, \sigma_{\Gamma_k},
\end{equation}
where $\sigma_{\Gamma_k}$ is the (fractional) 
theoretical uncertainty on partial width 
$\Gamma_k$.  The normalized derivatives are given analytically by
\begin{equation}
  \frac{\Gamma_k}{{\rm BR}_i}
  \frac{\partial {\rm BR}_{i}}{\partial \Gamma_k} = \left\{ 
  \begin{array}{ll}
  -{\rm BR}_k & \ {\rm for} \ i \neq k \\ 
  (1 - {\rm BR}_k) & \ {\rm for} \ i = k.
  \end{array} \right.
  \label{eq:dBRdGamma}
\end{equation}

\subsubsection{Cross section for $e^+e^- \to \nu \bar\nu H$}

At 1000~GeV centre-of-mass energy the production cross section for
$e^+e^- \to \nu \bar \nu H$ is dominated by $WW$ fusion, with a
subleading contribution from $e^+e^- \to ZH$ with $Z \to \nu \bar
\nu$.  The SM cross section is currently known including the one-loop
electroweak radiative
corrections~\cite{Belanger,Dittmaier1,Dittmaier2}.  We take the
remaining theoretical uncertainty in the cross section to be $0.5\%$;
see Appendix~\ref{app:xsec} for details.  This cross section
uncertainty is included in the $\chi^2$ calculation in the same way as
the luminosity uncertainty discussed earlier, completely correlated 
among the four rate measurements.

\subsubsection{$\alpha_s$, $\overline{m}_b(M_b)$, and $\overline{m}_c(M_c)$}

The most important sources of parametric uncertainties in the SM Higgs 
coupling calculations are the strong coupling $\alpha_s$ and the bottom
and charm quark masses.  The input values used in our analysis are 
summarized in Table~\ref{tab:paraminputs}.
For $\alpha_s$ we use the current world average from the Particle
Data Group~\cite{PDG}.
For the bottom and charm quark masses we use the values from fits to the 
kinematic moments in inclusive semileptonic $B$ meson decays; details are
given in Appendix~\ref{app:mbmc}.  We make the approximation that the
$\overline {\rm MS}$ quark mass $\overline{m_q}(M_q)$ evaluated at the 
corresponding quark pole mass $M_q$ (used by {\tt HDECAY})
is the same as the mass $\overline{m_q}(\overline{m_q})$ evaluated at its own
running mass (extracted from the $B$ decay fits).

\begin{table}
\begin{tabular}{cccc}
\hline \hline
Parameter & Value & Percent uncertainty & Source \\
\hline
$\alpha_s(m_Z)$ & $0.1185 \pm 0.0020$ & 1.7\% & \cite{PDG} \\
$\overline{m_b}(M_b)$ & $4.20 \pm 0.04$ GeV & 0.95\% & \cite{Buchmuller} \\
$\overline{m_c}(M_c)$ & $1.224 \pm 0.057$ GeV & 4.7\% & \cite{HoangBsl} \\
\hline\hline
\end{tabular}
\caption{Central values and uncertainties of 
$\alpha_s$ and the $\overline{\rm MS}$
bottom and charm quark masses $\overline{m_b}(M_b)$ and $\overline{m_c}(M_c)$
used in the analysis.}
\label{tab:paraminputs}
\end{table}

We implement the parametric uncertainties in the $\chi^2$ calculation
as follows.  Again, all uncertainties are fractional.
The correlated uncertainty $c_i^{x_j}$ in branching ratio $i$
due to the parametric uncertainty in 
$x_j \in \{ \alpha_s(m_Z), \overline{m_b}(M_b), \overline{m_c}(M_c) \}$
is given by
\begin{equation}
  c_i^{x_j} 
  = \frac{x_j}{{\rm BR}_i} 
  \frac{\partial {\rm BR}_i}{\partial x_j} \, \sigma_{x_j}
  = \sum_{k=1}^n 
  \left[
  \frac{\Gamma_k}{{\rm BR}_i} \frac{ \partial {\rm BR}_i}{\partial \Gamma_k} 
  \right]
  \left[
  \frac{x_j}{\Gamma_k} \frac{\partial \Gamma_k}{\partial x_j} \right] \,
  \sigma_{x_j},
\end{equation}
where $\sigma_{x_j}$ is the (fractional) uncertainty on the parameter $x_j$.
The second equality was obtained using the chain rule; the first
term inside the sum was given explicitly in Eq.~(\ref{eq:dBRdGamma}).

The calculation thus consists of evaluating the derivatives 
$\partial \Gamma_k / \partial x_j$ while holding constant all the parameters 
other than $x_j$.  In principle, this can be done by 
varying each of the inputs $x_j$ in {\tt HDECAY} and reading off the 
resulting variation in $\Gamma_k$.  This is complicated by the fact that 
{\tt HDECAY} takes as inputs not the running masses $\overline{m_b}(M_b)$
and $\overline{m_c}(M_c)$, but the pole masses $M_b$ and $M_c$ themselves.
Varying $M_b$ while holding $M_c$ constant is not the same as varying 
$\overline{m_b}(M_b)$ while holding $\overline{m_c}(M_c)$ constant; 
likewise, varying $\alpha_s(m_Z)$ while holding $M_b$ and $M_c$ constant
results in a variation of $\overline{m_b}(M_b)$ and $\overline{m_c}(M_c)$.

To deal with this, we use the chain rule and next-to-leading-order (NLO)
approximations.  We start by writing the quark pole masses as functions of
our desired input parameters: $M_b = M_b(\overline{m_b}, \alpha_s)$ and
$M_c = M_c(\overline{m_b}, \overline{m_c}, \alpha_s)$, 
where the running quark masses are understood to be evaluated at their 
corresponding pole mass scales and $\alpha_s \equiv \alpha_s(m_Z)$.
Any Higgs partial width $\Gamma$ can then be expressed as
$\Gamma = \Gamma(M_b(\overline{m_b}, \alpha_s), 
M_c(\overline{m_b}, \overline{m_c}, \alpha_s), \alpha_s)$.
The chain rule then gives
\begin{eqnarray}
   \left. \frac{\partial \Gamma}{\partial \overline{m_b}} 
       \right|_{\overline{m_c},\alpha_s}
   &=& \left. \frac{\partial \Gamma}{\partial M_b} \right|_{M_c,\alpha_s}
       \left. \frac{\partial M_b}{\partial \overline{m_b}} 
           \right|_{\alpha_s} 
     + \left. \frac{\partial \Gamma}{\partial M_c} \right|_{M_b,\alpha_s}
       \left. \frac{\partial M_c}{\partial \overline{m_b}}
           \right|_{\overline{m_c},\alpha_s} \\ \nonumber
   \left. \frac{\partial \Gamma}{\partial \overline{m_c}} 
       \right|_{\overline{m_b},\alpha_s}
   &=& \left. \frac{\partial \Gamma}{\partial M_c} \right|_{M_b,\alpha_s}
       \left. \frac{\partial M_c}{\partial \overline{m_c}} 
           \right|_{\overline{m_b},\alpha_s} \\ \nonumber
   \left. \frac{\partial \Gamma}{\partial \alpha_s} 
       \right|_{\overline{m_b},\overline{m_c}}
   &=& \left. \frac{\partial \Gamma}{\partial \alpha_s} \right|_{M_b,M_c}
     + \left. \frac{\partial \Gamma}{\partial M_b} \right|_{M_c,\alpha_s}
       \left. \frac{\partial M_b}{\partial \alpha_s} \right|_{\overline{m_b}}
     + \left. \frac{\partial \Gamma}{\partial M_c} \right|_{M_b,\alpha_s}
       \left. \frac{\partial M_c}{\partial \alpha_s} 
           \right|_{\overline{m_b},\overline{m_c}}.
\end{eqnarray}

{\tt HDECAY} can be used in a straightforward way to numerically
evaluate $[\partial \Gamma/\partial M_b]_{M_c,\alpha_s}$, $[\partial
\Gamma/\partial M_c]_{M_b,\alpha_s}$, and $[\partial \Gamma/\partial
\alpha_s]_{M_b,M_c}$.  We did this by varying each of the input
parameters $M_b$, $M_c$, and $\alpha_s$ about its central value over a
range of approximately twice its uncertainty as given in
Table~\ref{tab:paraminputs}.  We also checked explicitly that the
dependence of each of the partial widths was linear under these
(small) variations.

We evaluate the remaining derivatives analytically as follows.
The relation between the pole mass $M_q$ and running mass 
$\overline{m_q} \equiv \overline{m_q}(M_q)$ at NLO is 
\begin{equation}
  M_q = \overline{m_q} \left[ 1 + \frac{4}{3\pi} \alpha_s(M_q) \right].
  \label{eq:Mqas}
\end{equation} 
Differentiating this with respect to $\overline{m_q}$ yields
\begin{equation}
  \frac{\partial M_q}{\partial \overline{m_q}} 
  = 1 + \frac{4}{3\pi} \alpha_s(M_q) 
  + \overline{m_q} \frac{4}{3\pi} 
  \frac{\partial \alpha_s(M_q)}{\partial \overline{m_q}}.
\end{equation}
The derivative of $\alpha_s$ can be found from the NLO formula
\begin{equation}
  \alpha_s(M_q) 
  = \frac{\alpha_s(\mu)}{1 + (b/2\pi) \alpha_s(\mu) \log(M_q/\mu)},
  \label{eq:runningals}
\end{equation}
where $b = 11 - 2 N_f / 3$ and $N_f$ is the number of active quark flavours
in the running of $\alpha_s$ between the scales $\mu$ and $M_q$.
In particular, 
$\partial \alpha_s(M_q) / \partial \overline{m_q} = \mathcal{O}(\alpha_s^2)$;
working to NLO we will neglect this term.
The derivative of interest then becomes
\begin{equation}
  \frac{\partial M_q}{\partial \overline{m_q}} 
  \simeq 1 + \frac{4}{3} \alpha_s(M_q) = \frac{M_q}{\overline{m_q}},
\end{equation}
i.e., the normalized derivative 
$(\overline{m_q}/M_q)(\partial M_q/\partial \overline{m_q}) = 1$ at NLO.
A similar calculation yields 
$\partial M_c/\partial \overline{m_b} = 0$ at NLO.

The derivatives of the quark pole masses with respect to $\alpha_s$ are 
evaluated as follows.  From Eq.~(\ref{eq:Mqas}) we have, to NLO,
\begin{equation}
  \frac{\partial M_q}{\partial \alpha_s(m_Z)} 
  = \overline{m_q} \frac{4}{3\pi} 
  \frac{\partial \alpha_s(M_q)}{\partial \alpha_s(m_Z)}.
  \label{eq:dMqdals}
\end{equation}
Equation~(\ref{eq:runningals}) can be used to evaluate the $\alpha_s$
derivative at NLO,\footnote{This formula automatically takes into account
the change in the number of quark flavours in the running at the $b$ 
threshold.}
\begin{equation}
  \frac{\partial \alpha_s(M_q)}{\partial \alpha_s(m_Z)} 
  = \left[ \frac{\alpha_s(M_q)}{\alpha_s(m_Z)} \right]^2.
  \label{eq:dalsdals}
\end{equation}
Combining Eqs.~(\ref{eq:dMqdals}) and (\ref{eq:dalsdals}) we find for the
normalized derivative,
\begin{equation}
  \frac{\alpha_s(m_Z)}{M_q} \frac{\partial M_q}{\partial \alpha_s(m_Z)}
  = \frac{\overline{m_q}}{M_q} \frac{4}{3\pi} \alpha_s(M_q)
  \frac{\alpha_s(M_q)}{\alpha_s(m_Z)}
  \simeq \frac{4}{3\pi} \alpha_s(M_q),
\end{equation}
where in the last step we keep only terms up to NLO.

The dependences of the SM Higgs partial widths $\Gamma_i$ on variations in the
inputs $x_j \in \{ \alpha_s, \overline{m_b}(M_b), \overline{m_c}(M_c) \}$
are summarized in Table~\ref{tab:normderivs} in the form of normalized 
derivatives, $(x_j/\Gamma_i) (\partial \Gamma_i / \partial x_j)$, evaluated
at the central values of the inputs $x_j$.  For example, for $m_H = 120$ GeV,
a 1\% increase in $\overline{m_b}(M_b)$ leads to a 2.6\% increase 
in $\Gamma_{b \bar b}$, while a 1\% increase in $\alpha_s(m_Z)$ leads
to a 1.2\% decrease in $\Gamma_{b \bar b}$.
The dependence of the normalized derivatives on the Higgs mass is not very
strong.  We again use linear interpolation to find the appropriate 
normalized derivatives for Higgs mass values between those given in 
Table~\ref{tab:normderivs}.  Finally, we note that the Higgs partial widths
to $WW$, $ZZ$, and lepton pairs, and the production cross section for
$e^+e^- \to \nu \bar \nu H$, do not depend on 
$\alpha_s$, $\overline{m_b}(M_b)$, or $\overline{m_c}(M_c)$.

\begin{table}
\begin{tabular}{c|ccc|ccc|ccc}
\hline \hline
 \multicolumn{10}{c}{Normalized
 derivatives of Higgs partial widths} \\ 
\hline
 &\multicolumn{3}{c|}{$\alpha_s(m_Z)$} 
 &\multicolumn{3}{c|}{$\overline{m_b}(M_b)$} 
 &\multicolumn{3}{c}{$\overline{m_c}(M_c)$} \\ 
\hline
 $m_H$ &120 GeV&140 GeV&160 GeV&
        120 GeV&140 GeV&160 GeV&
        120 GeV&140 GeV&160 GeV \\ \hline
 $\Gamma_{b\bar{b}}$
&    $-1.177$
&    $-1.217$
&    $-1.249$
&     2.565
&     2.567
&     2.568
&     0.000
&     0.000
&     0.000
 \\ \hline
 $\Gamma_{c\bar{c}}$
&    $-4.361$
&    $-4.400$
&    $-4.432$
&    $-0.083$
&    $-0.084$
&    $-0.084$
&     3.191
&     3.192
&     3.192
 \\ \hline
 $\Gamma_{gg}$
&     2.277
&     2.221
&     2.175
&    $-0.114$
&    $-0.112$
&    $-0.104$
&    $-0.039$
&    $-0.032$
&    $-0.027$
 \\ \hline
 $\Gamma_{\gamma\gamma}$
&     0.002
&     0.002
&     0.001
&     0.010
&     0.008
&     0.005
&     0.012
&     0.009
&     0.005
 \\ 
\hline \hline
\end{tabular}
\caption{Normalized derivatives, $(x/\Gamma) (\partial \Gamma / \partial x)$, 
of SM Higgs partial widths with respect to the parameters $\alpha_s(m_Z)$, 
$\overline{m_b}(M_b)$, and $\overline{m_c}(M_c)$ for $m_H = 120$,
140, and 160 GeV.  See text for details.  The central values of the 
three input parameters were taken from Table~\ref{tab:paraminputs}.}
\label{tab:normderivs}
\end{table}

The general pattern of the normalized derivatives in Table~\ref{tab:normderivs}
can be understood as follows.
The $H \to q \bar q$ partial widths are proportional to 
$\overline{m}_q^2(m_H)$ [see Eq.~(\ref{eq:GH2qq})].
Neglecting QCD running of the quark mass, the normalized derivative of 
$\Gamma_{q \bar q}$ with respect to $\overline{m_q}$ should thus be equal
to 2.  The values in Table~\ref{tab:normderivs} are actually somewhat 
larger than 2 because of the effect of QCD running, which 
causes $\overline{m_q}(\mu)$ to decrease as the scale $\mu$ increases.
Raising $\overline{m_q}(M_q)$
by a small amount reduces the range of scales (from $M_q$ to $m_H$) over
which the QCD running is applied,
so that $\overline{m_q}(m_H)$ is even larger than it would be just from
the increase in $\overline{m_q}(M_q)$ without the effect of the running.
This effect is more pronounced in $\Gamma_{c \bar c}$ than in 
$\Gamma_{b \bar b}$ because $\alpha_s$ is larger at the charm quark mass 
scale than at the bottom mass scale, resulting in faster running for 
$\overline{m_c}$ near the scale $M_c$.
Similarly, increasing $\alpha_s(m_Z)$ strengthens the QCD running of 
$\overline{m_q}$, resulting in a reduction of $\Gamma_{q \bar q}$.  
Changing $\alpha_s$ also affects the QCD corrections to $\Gamma_{q \bar q}$;
however, the dominant effect is absorbed already into the running quark mass.

The slight dependence of $\Gamma_{c \bar c}$ on $\overline{m_b}(M_b)$
is due to the effect of the $b$ threshold on the running of
$\overline{m_c}$.  Adding the $b$ quark to the renormalization group
equations slows down the running of $\overline{m_c}$, so that
increasing the $b$ mass slightly reduces the high-scale value of
$\overline{m_c}$ and thus reduces $\Gamma_{c \bar c}$.  Note that by
working to NLO in evaluating the derivatives above, we have neglected
contributions to the $\overline{m_b}$ dependence of $\Gamma_{c \bar
c}$ of the same order as those that give the dependence shown in
Table~\ref{tab:normderivs}; a higher order treatment would thus give
different results.  However, the numerical effect of these terms is
negligible, so we do not attempt a more precise evaluation here.

$\Gamma_{gg}$ is proportional to $\alpha_s^2$ at leading order.  
NLO QCD corrections
increase $\Gamma_{gg}$ rather significantly, leading to a normalized 
derivative somewhat larger than 2.  $\Gamma_{gg}$ is dominated in the SM
by the top quark loop, so that the effect of varying $\overline{m_b}(M_b)$
and $\overline{m_c}(M_c)$ is small.  The anticorrelation in this case is
due to destructive interference between the real parts of the $b$ and $c$
quark loop amplitudes and the top quark loop amplitude.

Finally, the similar sensitivity of $\Gamma_{\gamma\gamma}$ to 
$\overline{m_b}(M_b)$
and $\overline{m_c}(M_c)$ can be understood by noting that the bottom and
charm quark loops contribute mainly through their interference with the 
dominant top and $W$ boson loops, and that their contributions go like
$\overline{m_b}(m_H) Q_b^2$ and $\overline{m_c}(m_H) Q_c^2$, respectively,
while $\overline{m_b}(\mu)/\overline{m_c}(\mu) 
\simeq 0.235 \pm 0.012$~\cite{Buchmuller} and $Q_c^2 = 4 Q_b^2$ so that 
the two contributions are actually comparable.

\section{Results}
\label{sec:results}

\begin{figure}
\resizebox{0.8\textwidth}{!}{\rotatebox{270}{\includegraphics{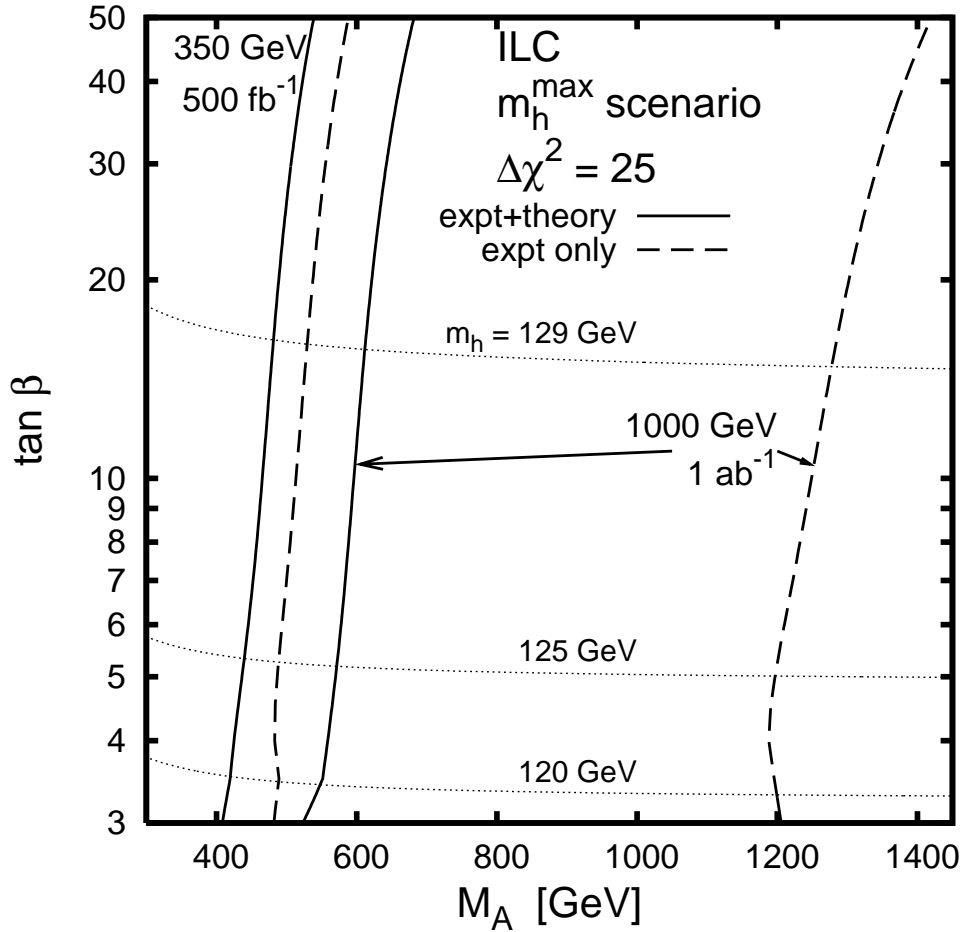}}}
\caption{Contours of $\Delta \chi^2 = 25$ from the experimental uncertainties
only (dashed lines) and including all theoretical and parametric 
uncertainties (solid lines).
The pair of contours on the left are for Phase 1 and the pair on the right
are for Phase 2.}
\label{fig:maincurves}
\end{figure}

Our main results are summarized in Fig.~\ref{fig:maincurves}, 
where we plot 
$\Delta \chi^2 = 25$ contours for the deviations between the SM Higgs and 
the lighter CP-even Higgs in the $m_h^{\rm max}$ scenario of the MSSM.
To the left of these contours, the MSSM Higgs can be distinguished from the
SM Higgs at more than the 5$\sigma$ level.
The dashed curves show the reach including experimental uncertainties only,
while the solid curves show the effect of adding the theoretical 
and parametric uncertainties.

\begin{figure}
\resizebox{0.8\textwidth}{!}{\rotatebox{270}{\includegraphics{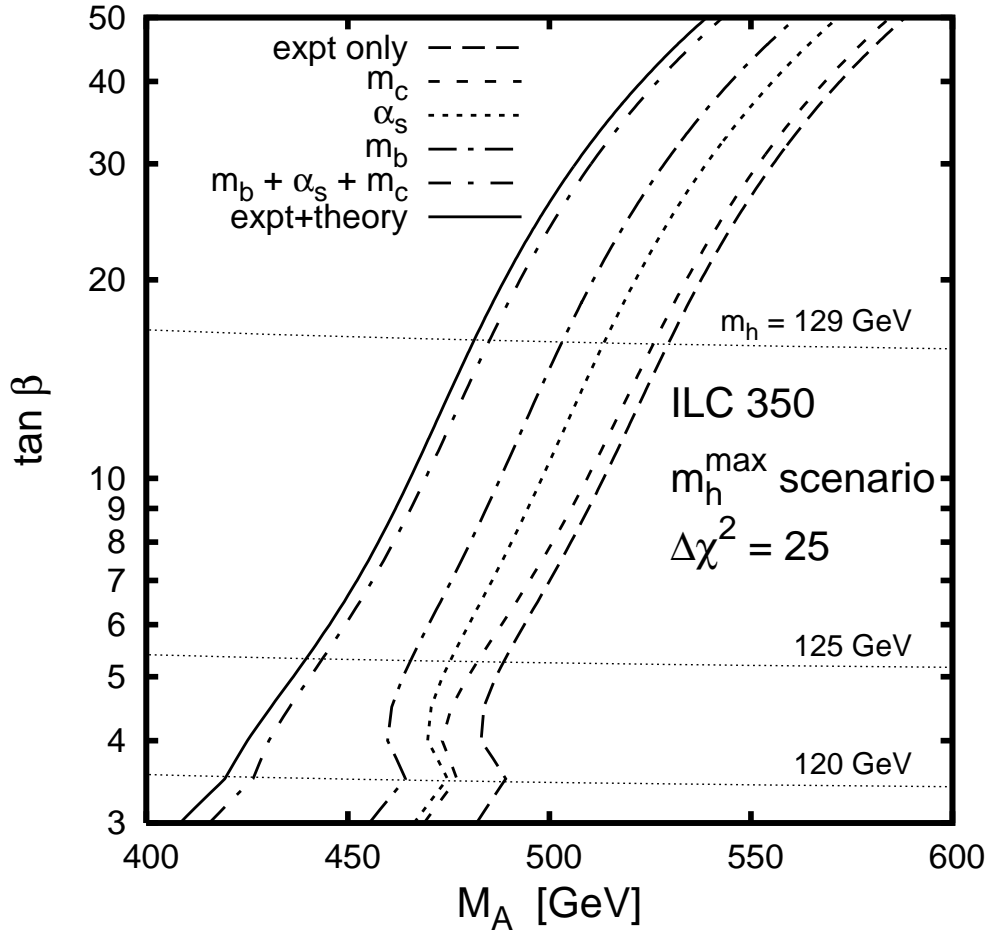}}}
\caption{Contributions of individual sources of parametric uncertainty
to the $\Delta \chi^2 = 25$ contours in Phase 1.}
\label{fig:phase1curvesPar}
\end{figure}

\begin{figure}
\resizebox{0.8\textwidth}{!}{\rotatebox{270}{\includegraphics{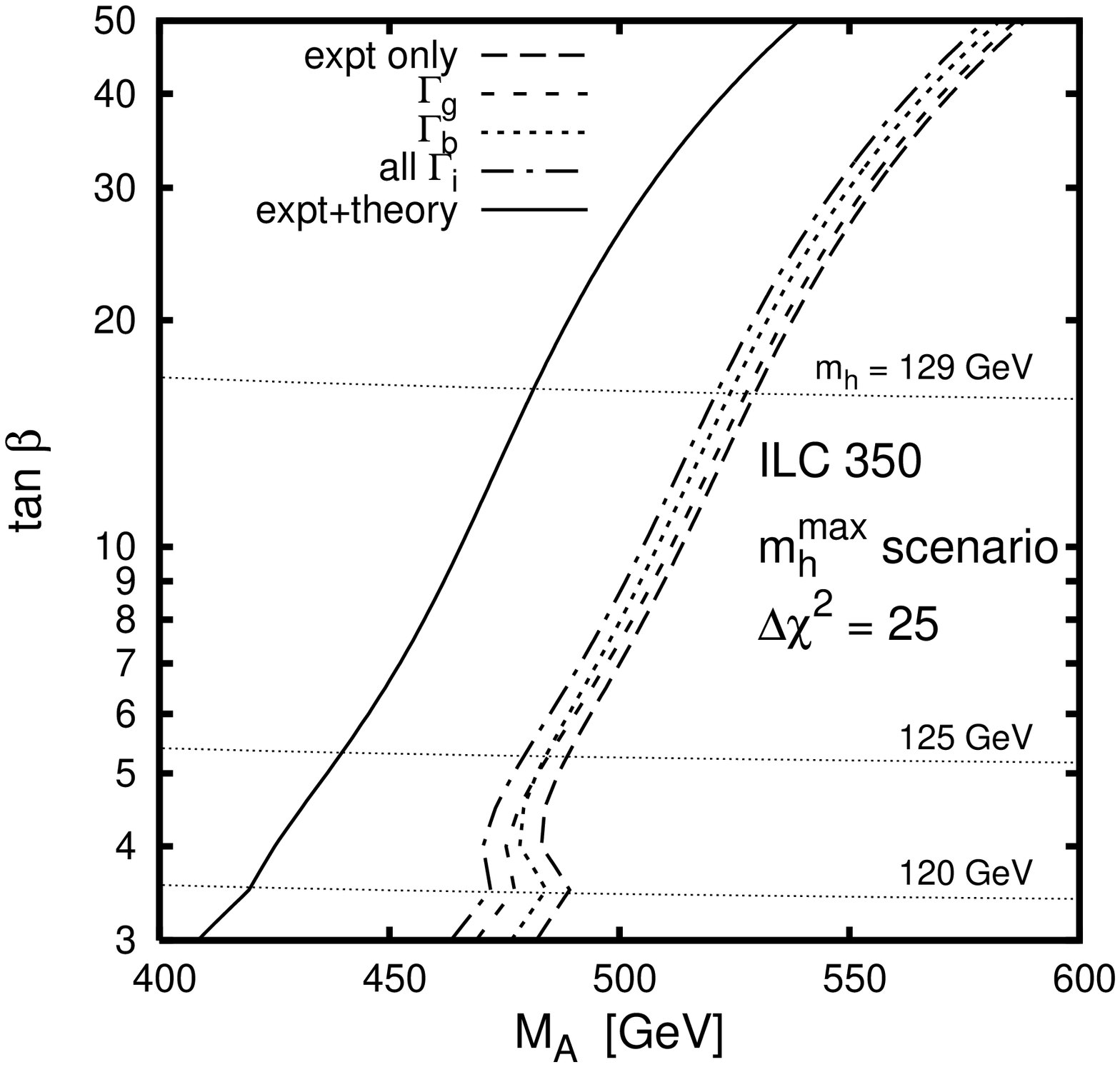}}}
\caption{Contributions of individual sources of theoretical uncertainty
to the $\Delta \chi^2 = 25$ contours in Phase 1.  The additional theory 
uncertainties not plotted separately have a negligible effect.}
\label{fig:phase1curvesThy}
\end{figure}

For the first phase of ILC running (the left pair of curves in
Fig.~\ref{fig:maincurves}; 500 fb$^{-1}$ at 350 GeV, experimental
uncertainties from Table~\ref{tab:p1uncert}), the impact of the
theoretical and parametric uncertainties is to reduce the reach in
$M_A$ by about 50~GeV, or roughly 10\%.  Details are shown in
Figs.~\ref{fig:phase1curvesPar} and \ref{fig:phase1curvesThy}, 
where we plot separately the contribution of each source of parametric and
theoretical uncertainty.  The bulk of the effect is due
to the parametric uncertainties in $\overline{m_b}(M_b)$ and
$\alpha_s(m_Z)$, as shown in Fig.~\ref{fig:phase1curvesPar}.  Smaller
contributions come from the theoretical uncertainty in $\Gamma_b$, and
for low $\tan \beta \lesssim 6$ from the parametric uncertainty in
$\overline{m_c}(M_c)$ and the theoretical uncertainty in $\Gamma_g$.
The change with $\tan\beta$ is partly due to the decrease
in $m_h$ with decreasing $\tan\beta$, as shown by the horizontal dotted 
lines in Figs.~\ref{fig:phase1curvesPar} and \ref{fig:phase1curvesThy}.

The relative importance of the different branching fraction measurements in
the first phase of ILC running is illustrated in Table~\ref{tab:phase1pulls}.  
\begin{table}
\begin{tabular}{ccccccc}
\hline \hline
\multicolumn{7}{c}{Phase 1 sample point: $M_A = 537.6$ GeV, $\tan\beta = 20$} \\ 
\hline
Observable & Shift & Expt uncert & Pull & Thy+param uncert & Total uncert & Pull \\
\hline
BR($b\bar{b}$) & 8.1\%     & 2.5\%  & 3.25    & 1.6\%  & 3.0\%  & 2.71 \\
BR($c\bar{c}$) & $-12.0\%$ & 13.2\% & $-0.90$ & 16.1\% & 20.8\% & $-0.57$ \\
BR($\tau\tau$) & 10.0\%    & 6.4\%  & 1.56    & 1.8\%  & 6.6\%  & 1.51 \\
BR($WW$)       & $-11.6\%$ & 3.9\%  & $-2.96$ & 1.8\%  & 4.3\%  & $-2.68$ \\ 
BR($gg$)       & $-14.7\%$ & 9.4\%  & $-1.56$ & 5.8\%  & 11.1\% & $-1.33$ \\
\hline \hline
\end{tabular}
\caption{Contributions to the $\Delta \chi^2$ at a Phase 1 sample point 
lying on the $\Delta \chi^2 = 25$ contour including only experimental 
uncertainties in Figs.~\ref{fig:phase1curvesPar} and 
\ref{fig:phase1curvesThy}.  
Shown are the fractional deviations in Higgs branching ratios 
in the $m_h^{\rm max}$ scenario compared to the SM, the experimental 
uncertainties for the corresponding Higgs mass $m_h = 129.2$~GeV
taken from a linear interpolation in 
Higgs mass between the values in Table~\ref{tab:p1uncert},
the resulting ``pulls,'' i.e., the branching ratio deviations normalized
by their experimental uncertainties, the combined theoretical and parametric
uncertainties on the observables, the total uncertainty, and the resulting
pull from the total uncertainty.  
Not shown are the off-diagonal elements of the correlation matrix
due to the theoretical and parametric uncertainties.}
\label{tab:phase1pulls}
\end{table}
We list the fractional deviations of the MSSM Higgs branching ratios from 
their SM values at the ``Phase 1 sample point'' 
$M_A = 537.6$~GeV and $\tan\beta = 20$, 
which lies on the $\Delta \chi^2 = 25$ contour from experimental 
uncertainties only and yields $m_h = 129.2$~GeV.
The relevant branching fractions deviate by roughly 10\% from their SM
values at this sample point.  The biggest contributions to the 
$\Delta \chi^2$ come from the best-measured branching fractions, 
BR($b \bar b$) and BR($WW$).  Including the theoretical and parametric
uncertainties moderately degrades the precision in all the channels, 
reducing the $\Delta \chi^2$ from 25 down to 17.4 at the sample point.
This $\Delta \chi^2$ is slightly worse than the value of 18.9 that would 
be obtained by summing the squares of the pulls in Table~\ref{tab:phase1pulls};
this is due to the effect of the off-diagonal elements in the correlation
matrix.  The effects of the most important parametric uncertainties, in 
$\overline{m_b}(M_b)$ and $\alpha_s(m_Z)$, are anticorrelated between 
BR($b \bar b$) and BR($WW$) and thus can mimic the anticorrelated shift in
these observables that occurs in the MSSM.

\begin{figure}
\resizebox{0.8\textwidth}{!}{\rotatebox{270}{\includegraphics{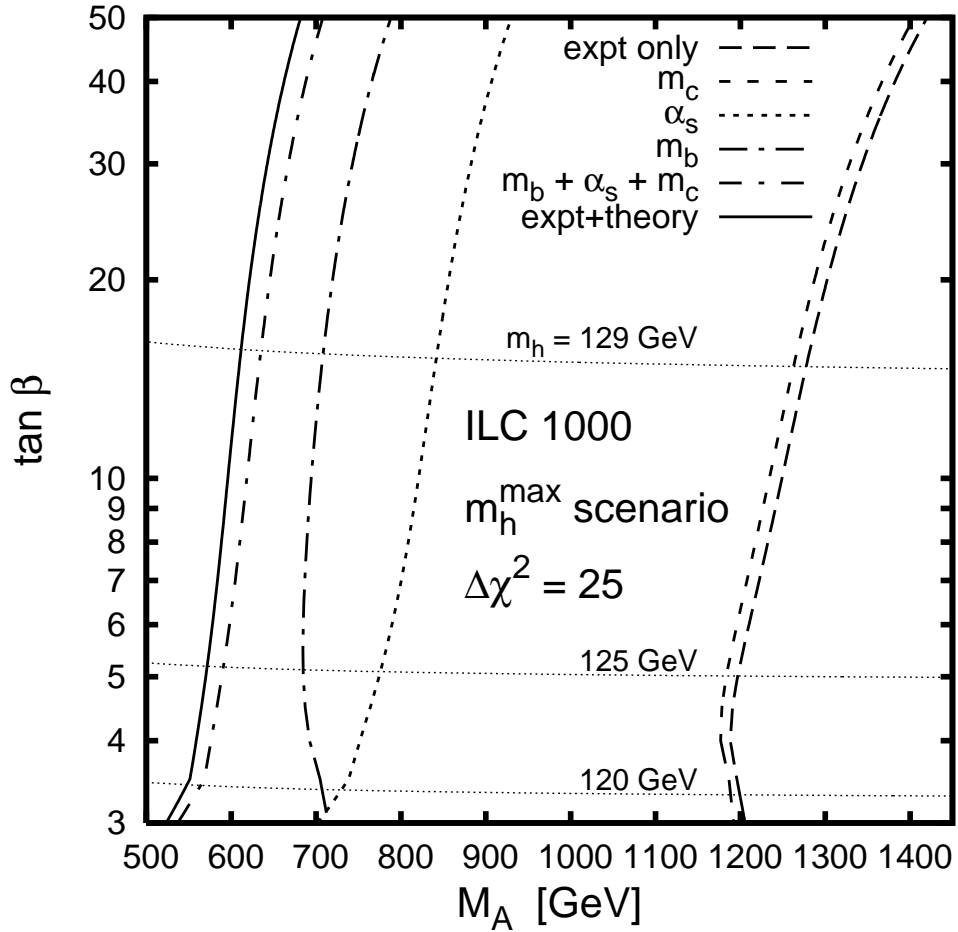}}}
\caption{Contributions of individual sources of parametric uncertainty
to the $\Delta \chi^2 = 25$ contours in Phase 2.}
\label{fig:phase2curvesPar}
\end{figure}

\begin{figure}
\resizebox{0.8\textwidth}{!}{\rotatebox{270}{\includegraphics{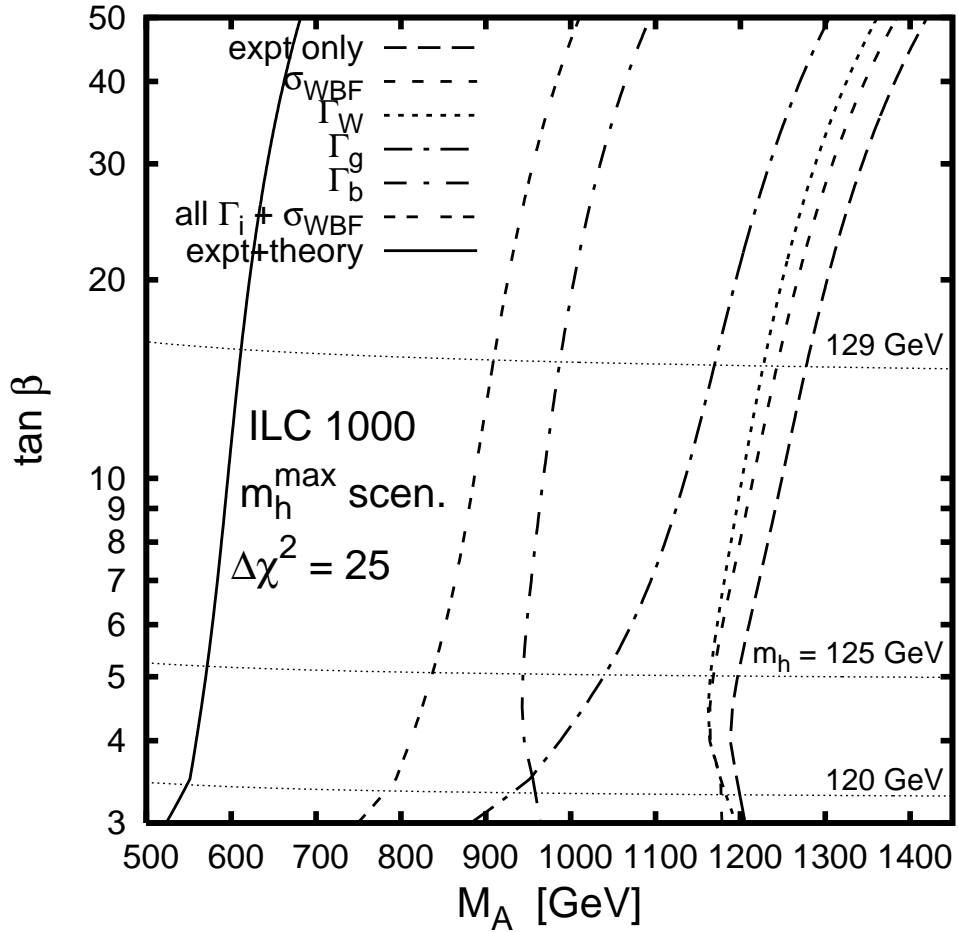}}}
\caption{Contributions of individual sources of theoretical uncertainty
to the $\Delta \chi^2 = 25$ contours in Phase 2.  The additional theory 
uncertainties not plotted separately have a negligible effect.}
\label{fig:phase2curvesThy}
\end{figure}

For the second phase of ILC running (the right set of curves in
Fig.~\ref{fig:maincurves}; combining measurements from 1000 fb$^{-1}$
at 1000 GeV and 500 fb$^{-1}$ at 350 GeV, experimental uncertainties
from Tables~\ref{tab:p1uncert} and \ref{tab:p2uncert}), the
experimental uncertainties are considerably smaller.  The theoretical
and parametric uncertainties thus have a much more significant impact:
their effect is to reduce the reach in $M_A$ by about a factor of two,
from about 1200--1400~GeV down to about 525--700~GeV.  Details are
shown in Figs.~\ref{fig:phase2curvesPar} and \ref{fig:phase2curvesThy}.  
The dominant contributions by far are the
parametric uncertainties in $\overline{m_b}(M_b)$ and $\alpha_s(m_Z)$
(Fig.~\ref{fig:phase2curvesPar}), followed by the theoretical
uncertainty in $\Gamma_b$ and for low $\tan\beta \lesssim 4$ in
$\Gamma_g$ (Fig.~\ref{fig:phase2curvesThy}).

The impact of the parametric and theoretical uncertainties on the 
individual channels used in the Phase 2 $\Delta \chi^2$ calculation
is illustrated in Table~\ref{tab:phase2pulls}.
\begin{table}
\begin{tabular}{ccccccc}
\hline \hline
\multicolumn{7}{c}{Phase 2 sample point: $M_A = 1302.4$ GeV, $\tan\beta = 20$} \\ 
\hline
Observable & Shift & Expt uncert & Pull & Thy+param uncert & Total uncert & Pull \\
\hline
BR($b\bar{b}$) & 1.7\%    & 2.5\%  & 0.67    & 1.7\%  & 3.0\%  & 0.55 \\
BR($c\bar{c}$) & $-2.5\%$ & 13.3\% & $-0.19$ & 16.1\% & 20.8\% & $-0.12$ \\
BR($\tau\tau$) & 2.1\%    & 6.4\%  & 0.34    & 1.8\%  & 6.6\%  & 0.32 \\
BR($WW$)       & $-2.1\%$ & 3.9\%  & $-0.53$ & 1.8\%  & 4.3\%  & $-0.48$ \\ 
BR($gg$)       & $-4.6\%$ & 9.4\%  & $-0.48$ & 5.8\%  & 11.1\% & $-0.41$ \\
\hline
$\sigma \times {\rm BR}(b \bar{b})$     & 1.7\%    & 0.45\% & 3.72    & 1.7\% & 1.8\% & 0.93 \\
$\sigma \times {\rm BR}(WW)$            & $-2.1\%$ & 0.93\% & $-2.22$ & 1.9\% & 2.1\% & $-0.98$ \\ 
$\sigma \times {\rm BR}(gg)$            & $-4.6\%$ & 2.0\%  & $-2.32$ & 5.8\% & 6.2\% & $-0.74$ \\
$\sigma \times {\rm BR}(\gamma \gamma)$ & 0.27\%   & 5.5\%  & 0.05    & 1.9\% & 5.8\% & 0.05 \\
\hline \hline
\end{tabular}
\caption{Contributions to the $\Delta \chi^2$ at a Phase 2 sample point 
lying on the $\Delta \chi^2 = 25$ contour including only experimental 
uncertainties in Figs.~\ref{fig:phase2curvesPar}
and \ref{fig:phase2curvesThy}.  
Shown are the fractional deviations in Higgs observables
in the $m_h^{\rm max}$ scenario compared to the SM, the experimental 
uncertainties for the corresponding Higgs mass $m_h = 129.3$~GeV
taken from a linear interpolation in 
Higgs mass between the values in Tables~\ref{tab:p1uncert} 
and \ref{tab:p2uncert},
the resulting pulls, the combined theoretical and parametric
uncertainties on the observables, the total uncertainty, and the resulting
pull from the total uncertainty.  
Not shown are the off-diagonal elements of the correlation matrix
due to the theoretical and parametric uncertainties.}
\label{tab:phase2pulls}
\end{table}
We list the fractional deviations of the MSSM Higgs observables 
from their SM values at the ``Phase 2 sample point'' 
$M_A = 1302.4$~GeV and $\tan\beta = 20$,
which lies on the $\Delta \chi^2 = 25$ contour from experimental uncertainties 
only and yields $m_h = 129.3$~GeV.
The relevant observables deviate by roughly 2\% to 4\% from their
SM values at this sample point, except for $\sigma \times$BR($\gamma\gamma$) 
which is very close to its SM value.
Ignoring the parametric and theoretical uncertainties,
the biggest contribution to the $\Delta \chi^2$ comes from the best-measured
observable, $\sigma \times {\rm BR}(b \bar b)$, followed by 
$\sigma \times {\rm BR}(gg)$ and $\sigma \times {\rm BR}(WW)$.
The contribution of the branching fraction measurements from 
350~GeV running to the $\Delta \chi^2$ is small.

Including the theoretical and parametric uncertainties severely degrades
the precision of the three most important measurements,
$\sigma \times {\rm BR}(b \bar b)$, $\sigma \times {\rm BR}(WW)$, 
and $\sigma \times {\rm BR}(gg)$, reducing the $\Delta \chi^2$ 
dramatically from 25 down to 1.7 at the sample point.  
This $\Delta \chi^2$ is somewhat
worse than the value of 3.2 that would be obtained by summing the squares
of the pulls in Table~\ref{tab:phase2pulls}; this is again due to the 
effect of the off-diagonal elements in the correlation matrix, including
the correlations between the Phase 1 branching ratio measurements and the 
Phase 2 rate measurements.

\section{Discussion and conclusions}
\label{sec:conclusions}

We studied the impact of the theoretical and parametric uncertainties
in the SM predictions for Higgs production and decay on the ability of
the ILC to reveal deviations from the SM Higgs.
To quantify the impact of the theoretical and parametric uncertainties, we 
compared the SM predictions for Higgs observables to those in the 
$m_h^{\rm max}$ benchmark scenario of the MSSM.  We plotted 
$\Delta \chi^2 = 25$ contours for the deviations between the SM and MSSM 
Higgs observables both with and without the theory uncertainties, given the 
expected precisions on Higgs measurements at a future ILC.
We found that the impact of the theoretical and parametric uncertainties is 
moderate in the first phase of ILC data-taking (500 fb$^{-1}$ at 350 GeV 
centre-of-mass energy), reducing the reach in the CP-odd MSSM Higgs
mass $M_A$ by about 10\% to $\sim 500$~GeV, 
while in the second phase (1000 fb$^{-1}$ at 1000 
GeV) these uncertainties are larger than the experimental uncertainties and
reduce the reach in $M_A$ by about a factor of two, from $\sim 1200$ down
to $\sim 600$~GeV.  The bulk of the effect comes from the parametric 
uncertainties in $m_b$ and $\alpha_s$, followed by the theoretical 
uncertainty in $\Gamma_b$.  The theoretical uncertainty in $\Gamma_g$
is also important for lower Higgs masses below about 120--125~GeV.

The single most important source of theoretical or parametric
uncertainty is the bottom quark mass.  In our analysis we used the
measurement of $m_b$ from a global fit to inclusive semileptonic $B$
meson decay spectra~\cite{Buchmuller} (see Appendix~\ref{app:mbmc} for
details).  This is currently the best measurement of $m_b$; its
uncertainty of 0.95\% is dominated by the theoretical uncertainties in
the fit.  The large impact of this parametric uncertainty on Higgs
observables is due to {\it (i)} the effective 2.6 power with which
$\overline{m_b}(M_b)$ enters $\Gamma_b$, {\it (ii)} the importance of
the $b \bar b$ final state in ILC Higgs measurements, and {\it (iii)}
the fact that the deviations of the MSSM Higgs couplings from their SM
values show up largely in the $b$ and $\tau$ sectors.  The impact of
the uncertainty in $m_b$ on the long-term ILC Higgs program highlights
the need for further theoretical work on the $m_b$ extraction from $B$
meson observables.

\begin{figure}
\resizebox{0.8\textwidth}{!}{\rotatebox{270}{\includegraphics{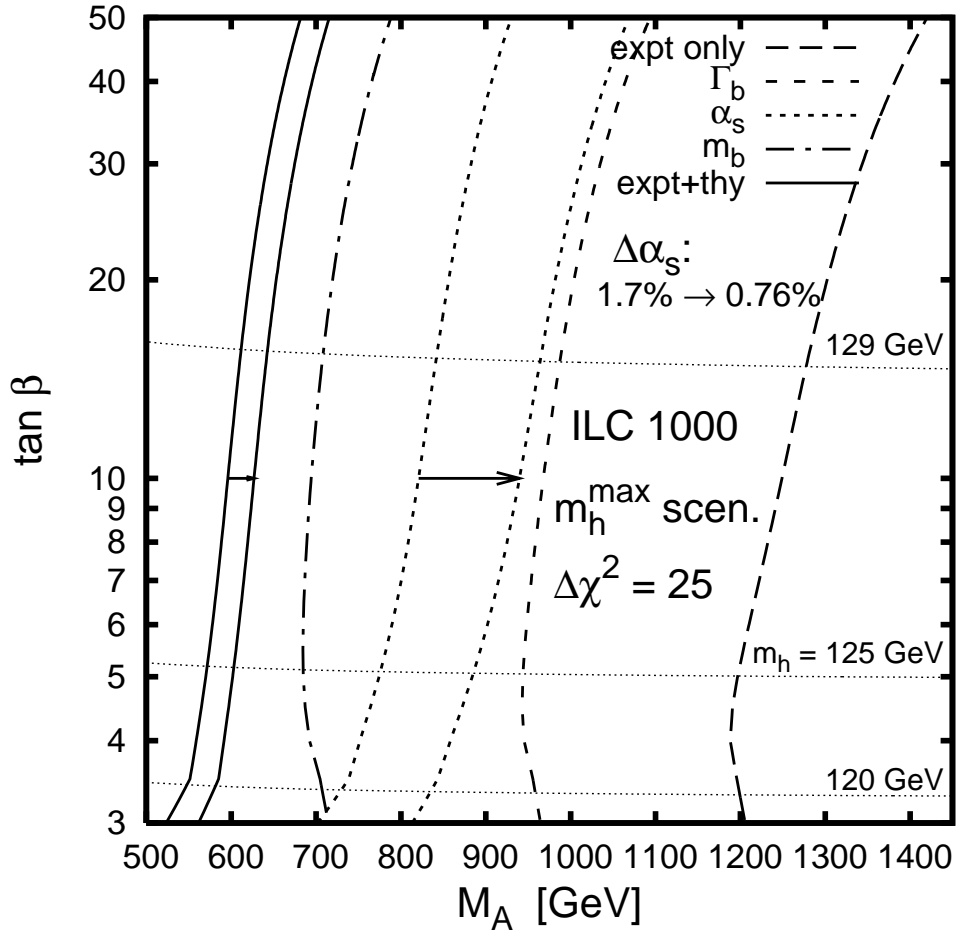}}}
\caption{Effect on the Phase 2 $\Delta \chi^2 = 25$ contours of improving 
the uncertainty on $\alpha_s(m_Z)$ from 1.7\% to 0.76\%, as projected for 
GigaZ in Ref.~\cite{TeslaTDR}.}
\label{fig:als}
\end{figure}

The second most important source of theoretical or parametric
uncertainty is $\alpha_s(m_Z)$.  In our analysis we used the
world-average PDG value~\cite{PDG} with an uncertainty of 0.0020 or
1.7\%.  This measurement is expected to be improved by at least a
factor of two at the ILC through precision measurements of event shape
observables, the cross section ratio $\sigma_{t \bar
t}/\sigma_{\mu^+\mu^-}$ above the $t \bar t$ threshold, and
$\Gamma_Z^{\rm had}/\Gamma_Z^{\rm lept}$ at the $Z$ pole (via the GigaZ
option)~\cite{TeslaTDR}.  We illustrate the effect of improving the
uncertainty on $\alpha_s(m_Z)$ to 0.0009 or 0.76\%~\cite{TeslaTDR} in
Fig.~\ref{fig:als}.
By itself this improvement has a relatively small effect because the
uncertainty is still dominated by $m_b$; however, if the $m_b$
extraction becomes significantly more precise, the improvement in
$\alpha_s$ will be valuable.

The measurement of BR($cc$) did not contribute significantly to the
$\Delta \chi^2$ because of its relatively poor experimental precision,
particularly at the higher Higgs masses $\sim 125$--130~GeV that
appear over most of the $m_h^{\rm max}$ parameter space.  Even if the
ILC measurement were to be improved, however, the parametric
uncertainties in $\overline{m_c}(M_c)$ and $\alpha_s(m_Z)$ lead to a
large uncertainty in the SM prediction for BR($cc$) -- 16\% at our
sample points -- limiting the usefulness of this mode for model
discrimination.  The charm mass is by far the dominant uncertainty
here, contributing 14.6\%, while $\alpha_s$ contributes 6.4\%.  Note
however that the theoretical improvements in $B$ meson decays required
to reduce the uncertainty in $m_b$ will also reduce the uncertainty in
$m_c$.

We also note here that the theoretical calculation of higher-order QCD
corrections to $H \to gg$ includes processes in which one of the
final-state gluons splits into a $c \bar c$ or $b \bar b$ pair.
Determining which part of these processes should be included in the $H
\to gg$ branching fraction and which in the $H \to c \bar c$ or $b
\bar b$ branching fractions will require interaction between the
theoretical and experimental studies.  The impact of this splitting on
the $b \bar b$ final state is small, but it could potentially shift
the numbers of events tagged as $c \bar c$ or $gg$ by tens of percent
(see the end of Appendix~\ref{sec:Hgg} for details).  We have not
attempted to address this question here.

To understand the effect of the correlated uncertainties, we recall that
for large $\tan\beta$, $\Gamma_b$ and $\Gamma_{\tau}$ exhibit the largest
deviations from their SM values, while $\Gamma_W$ approaches its SM value
very quickly with increasing $M_A$ [Eq.~(\ref{eq:deviations})].
The largest parametric and theoretical uncertainties appear in the hadronic
decay widths of the Higgs.  At the same time, the most precisely measured 
observables in Phase 2 of the ILC 
are $\sigma \times {\rm BR}(bb)$ and $\sigma \times {\rm BR}(gg)$, 
which suffer directly from the parametric and theoretical uncertainties,
and $\sigma \times {\rm BR}(WW)$, which is affected indirectly through
the Higgs total width.  Adding a high-precision measurement of 
$\sigma \times {\rm BR}(\tau\tau)$ would allow a tight constraint 
on the MSSM Higgs unpolluted by the dominant uncertainties.
This mode is affected by the parametric and theoretical uncertainties 
indirectly through the Higgs total width in the same way as 
$\sigma \times {\rm BR}(WW)$, so that the ratio of these two non-hadronic
modes is relatively clean.  Further, $\Gamma_{\tau}$ exhibits a large
deviation from its SM value in the MSSM [see Eq.~(\ref{eq:deviations})], 
leading to good potential sensitivity to the MSSM nature of the Higgs.  

\begin{figure}
\resizebox{0.8\textwidth}{!}{\rotatebox{270}{\includegraphics{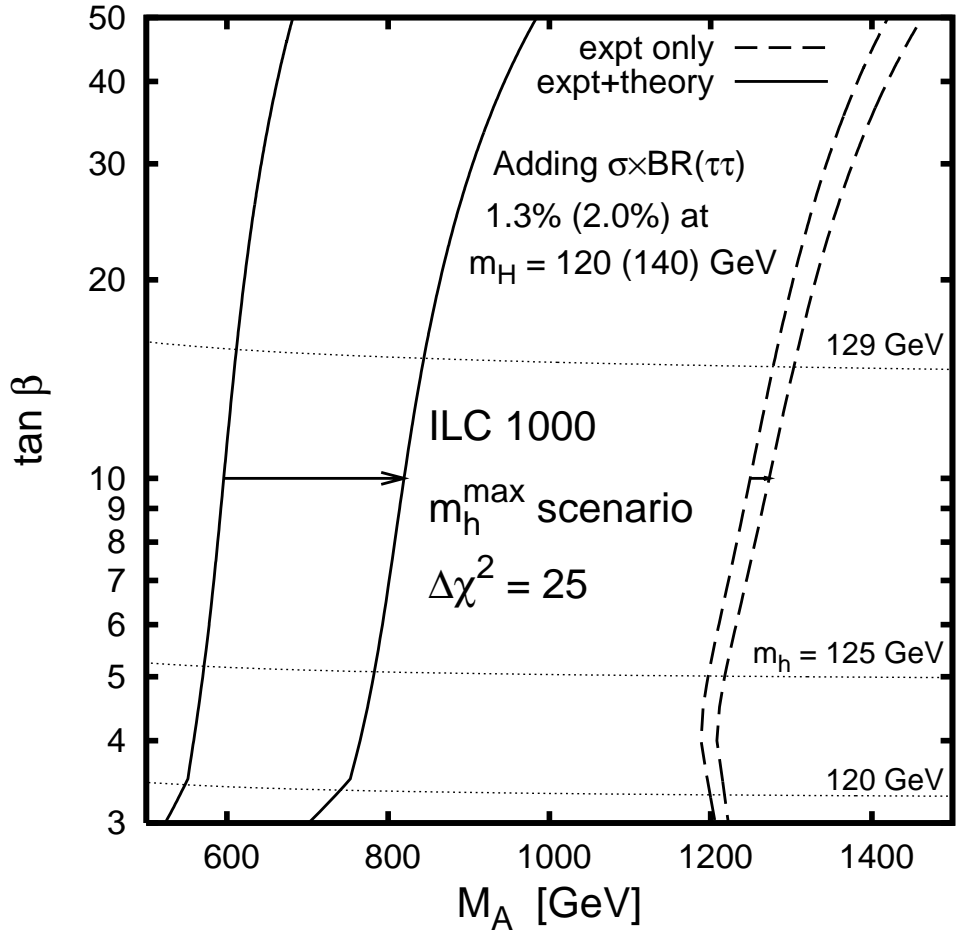}}}
\caption{Effect on the Phase 2 $\Delta \chi^2 = 25$ contours of adding 
a measurement of $\sigma \times {\rm BR}(\tau\tau)$ with a precision of
1.3\% (2.0\%) at $m_H = 120$ (140)~GeV.}
\label{fig:tau}
\end{figure}

To illustrate the potential improvement, we first note that for $m_H$
between 120 and 140~GeV, the experimental uncertainties in the $b \bar
b$, $WW$, and $gg$ final states improve by factors of 5--6, 4--5, and
3.5--5.5, respectively, between Phase 1 and Phase 2 (see
Tables~\ref{tab:p1uncert} and \ref{tab:p2uncert}).  We thus expect
that an improvement in the $\tau\tau$ final state in Phase 2 by a
factor of about four should be reasonable, and show the impact of
a measurement of $\sigma \times {\rm BR}(\tau\tau)$ with an
uncertainty of 1.3\% (2.0\%) at $m_H = 120$ (140)~GeV in Fig.~\ref{fig:tau}.
The effect of such a measurement on the reach in $M_A$ including only
experimental uncertainties is minor, because the precision assumed here 
is not particularly high compared to that of $\sigma \times {\rm BR}(bb)$
and $\sigma \times {\rm BR}(WW)$.
However, the impact on the reach including parametric and theoretical 
uncertainties is quite significant, improving the reach from $\sim 600$~GeV
to $\sim 800$~GeV.
A measurement of $\sigma \times {\rm BR}(\tau\tau)$ would require a 
different selection than used in Ref.~\cite{Barklow}, which 
required the visible energy in the event to add up 
to the Higgs mass; in the case of $H \to \tau\tau$, some energy is lost 
through the neutrinos in the $\tau$ decays.

As the experimental studies for the ILC Higgs measurements are
refined, we hope that the correlations among the experimental
uncertainties in different channels will be quantified.  In
particular, we expect the hadronic decay modes $H \to gg$, $b \bar b$ and 
$c \bar c$ to be correlated, because the separation of these three channels
relies on bottom and charm tagging.  Because the shifts in Higgs
observables in particular models are correlated due to the model
structure, we expect the correlations in the experimental
uncertainties to be important for evaluating model distinguishability.
Once these correlated uncertainties are available, including them in
a $\chi^2$ framework will be straightforward.

We used the $m_h^{\rm max}$ benchmark scenario of the MSSM to quantify
the impact of the theoretical and parametric uncertainties.  Our
analysis can be extended in a straightforward way to Higgs coupling
comparisons in other scenarios of the MSSM and to other beyond-the-SM
Higgs sectors.  Such a study would be interesting because the
characteristic deviations from the SM in different extended models may
appear in different channels than for the MSSM $m_h^{\rm max}$
scenario, so that the relative importance of the various Higgs
measurements will differ.  Further, the shifts in the observables
predicted in different extended Higgs models may pull in different
directions relative to the correlations in the uncertainties in the SM
predictions.

\begin{acknowledgments}
We are grateful to Tim Barklow and Klaus Desch for clarifying the experimental
studies of Higgs production and decay at the ILC, to Stefan Dittmaier for 
an enlightening discussion on the high-precision calculation of the 
$e^+e^- \to \nu \bar \nu H$ cross section, and to Thomas Becher, Andre
Hoang, Andreas Kronfeld and Frank Petriello for useful conversations on the 
bottom and charm quark mass uncertainties.
This work was supported by the Natural Sciences and Engineering
Research Council of Canada.
\end{acknowledgments}


\appendix

\section{Theoretical uncertainties in Higgs decay partial widths}
\label{app:thy}

The radiative corrections to Higgs decays have been reviewed in
Ref.~\cite{Spirareview}; we give here a brief sketch of the known
corrections.  The theory uncertainties used in our analysis were
summarized in Table~\ref{tab:thyuncert}.

\subsection{$H \to q \bar q$}

The QCD corrections to $H \to q \bar q$ are known up to three 
loops (N$^3$LO)~\cite{Braaten:1980yq,Gorishnii:1990zu,Chetyrkin:1995pd} for
massless final-state quarks.
A compact formula in the $\overline{\rm MS}$ scheme is given in
Ref.~\cite{Spirareview}:
\begin{equation}
  \Gamma(H \to q\bar q) = \frac{3 G_F m_H}{4\sqrt{2}\pi} \overline{m}_q^2(m_H)
  \left[ \Delta_{\rm QCD} + \Delta_t \right],
  \label{eq:GH2qq}
\end{equation}
where
\begin{eqnarray}
  \Delta_{\rm QCD} &=& 1 + 5.67 \frac{\alpha_s(m_H)}{\pi}
  + (35.94 - 1.36 N_F) \left( \frac{\alpha_s(m_H)}{\pi} \right)^2
  \nonumber \\
  & & + (164.14 - 25.77 N_F + 0.259 N_F^2) 
    \left( \frac{\alpha_s(m_H)}{\pi} \right)^3
  \nonumber \\
  &\simeq & 1 + 0.20 + 0.04 + 0.003
  \nonumber \\
  \Delta_t &=& \left( \frac{\alpha_s(m_H)}{\pi} \right)^2
  \left[ 1.57 - \frac{2}{3} \log \frac{m_H^2}{M_t^2}
    + \frac{1}{9} \log^2 \frac{\overline{m}_q^2(m_H)}{m_H^2} \right]
  \nonumber \\
  &\simeq & 0.02.
  \label{eq:DeltaHqq}
\end{eqnarray}
In this calculation, the $\overline {\rm MS}$ running quark masses 
$\overline{m_q}(M_q)$ evaluated at the quark pole mass $M_q$ are taken as 
a starting point 
and then run up to the Higgs mass scale using the standard QCD running in 
order to absorb large logarithms.  The coefficient functions for the 
running of the quark masses are known to four-loop 
accuracy~\cite{Chetyrkin:1997dh}, as is the beta-function for the
running of $\alpha_s$~\cite{vanRitbergen:1997va}.
We thus estimate the uncertainty in the partial width from the running
of $\overline{m}_q(m_H)$ to be well below the 0.01\% level.  

The perturbative expansion for $\Delta_{\rm QCD}$ in Eq.~(\ref{eq:DeltaHqq}) 
is well under control;
$\Delta_t$ appears at NNLO and is roughly the same size as the NNLO term
in $\Delta_{\rm QCD}$.  We thus estimate the uncertainty from uncalculated
higher-order QCD corrections to be smaller than the last calculated term
in each series: $\sim 0.1\%$ from $\Delta_{\rm QCD}$ and $\sim 1\%$ 
from $\Delta_t$.

We note here that an additional contribution to $H \to q \bar q$ can come 
from $H \to gg^*$ with the off-shell gluon splitting to $q \bar q$; we have
\emph{not} included these effects in our calculations.  See 
Sec.~\ref{sec:Hgg} for further discussion.

For Higgs decays to heavy quarks near the threshold region, the quark
pole mass dependence must be included through the usual $\beta^3 = (1
- 4 M_q^2/m_H^2)^{3/2}$ factor.  The NLO QCD corrections for massive
quarks are known~\cite{Braaten:1980yq}, as is the $\mathcal{O}(N_f
\alpha_s^2)$ piece of the massive NNLO
corrections~\cite{Melnikov:1995yp}.  {\tt HDECAY} uses a linear
interpolation between the NLO massive and the N$^3$LO massless results
near the threshold region~\cite{Spirareview}.  For $m_H \sim 120$~GeV,
the quark mass effects decrease the $H \to b \bar b$ ($c \bar c$)
partial width by about 0.7\% (0.1\%).  We estimate the remaining
uncertainty from quark mass effects in the QCD corrections to be below
the 0.01\% level.

The electroweak corrections to the Higgs decay into fermion
pairs are known at 
one loop~\cite{Fleischer:1980ub,Bardin:1990zj}; a convenient approximation
formula is given in Refs.~\cite{Djouadi:1991uf,Spirareview}.
The QED corrections include a large logarithm $\log(m_H^2/M_q^2)$, which
can be absorbed into the running fermion mass as in QCD.
The electroweak corrections also include a top-mass-dependent piece which
increases the $H \to q \bar q$ partial width by about 2.2\% for decays 
to $c \bar c$ and by 0.3\% for decays to $b \bar b$ (the difference is
due to the vertex correction involving a top quark in the case of 
$b \bar b$, which partly cancels the universal correction).  In addition, 
the two- and three-loop QCD corrections to the 
top-mass-dependent piece of the electroweak correction are 
known~\cite{Kniehl:1994ju,Chetyrkin:1996wr}.  Their numerical size depends
on the scheme chosen to define the running $\alpha_s$ and top quark 
mass~\cite{Chetyrkin:1996wr}; a suitable choice gives a \emph{relative} 
correction to the top-mass-dependent piece of the NLO electroweak 
correction of 8\% (two-loop) and 1\% (three-loop).
The remaining electroweak corrections are comparable in size, i.e., at 
the percent level.  We thus estimate the remaining theoretical uncertainty 
from higher-order electroweak and mixed electroweak/QCD corrections to
be smaller by a loop factor $\sim \alpha/\pi$, i.e., at the level of 0.01\%.

For large Higgs masses, electroweak corrections due to the Higgs
self-coupling become relevant.  These were calculated in
Ref.~\cite{Ghinculov:1994se} and can increase $\Gamma_{q \bar q}$ by
about 2\% for $m_H \sim 1$ TeV.  For a 120 GeV Higgs, they increase
$\Gamma_{q \bar q}$ by about 0.2\% at NLO and by another $-0.002\%$ at
NNLO.  Any remaining uncertainty from higher orders in the Higgs
self-coupling is thus totally negligible in the intermediate Higgs
mass range.

We thus estimate the theory uncertainty in $\Gamma_{b \bar b}$ and 
$\Gamma_{c \bar c}$ to be at the 1\% level, primarily from the $\Delta_t$
piece of the QCD corrections.

\subsection{$H \to \ell^+ \ell^-$}

Higgs decays to leptons receive the same electroweak corrections as
the decays to quarks discussed in the previous section.  As mentioned
previously, we estimate that the theoretical uncertainty is at the 
level of two-loop electroweak corrections $\sim (\alpha/\pi)^2 \sim 0.01\%$.

\subsection{$H \to WW$, $ZZ$}

For Higgs masses below the $WW$ threshold,
decays with one or both gauge bosons off-shell are important,
with branching ratios for the doubly off-shell decays reaching the percent 
level for Higgs masses above about 100--110 GeV~\cite{Spirareview}.
Until recently, calculations for off-shell vector bosons were only available
at tree level, and radiative corrections to Higgs decays to $WW$ and 
$ZZ$ were known only in the narrow-width approximation. 
In this approximation, the one-loop electroweak 
corrections~\cite{Fleischer:1980ub,Kniehl:1990mq,Kniehl:1991xe,Bardin:1991dp}
amount to about 5\% or less in the intermediate mass 
range~\cite{Spirareview}.
Furthermore, the QCD corrections to the leading top-mass-enhanced
electroweak corrections of $\mathcal{O}(\alpha_s G_F m_t^2)$ 
have been calculated at two~\cite{Kniehl:1995tn,Kniehl:1995gj}
and three~\cite{Kniehl:1995br,Kniehl:1995at} loops; these mixed 
electroweak/QCD corrections amount to roughly 0.5\%.
{\tt HDECAY} takes into account both singly off-shell and doubly off-shell
decays at tree level but neglects these electroweak and mixed 
corrections~\cite{Spirareview}.

Recently the complete NLO strong and electroweak radiative corrections for 
$H \to WW/ZZ \to 4 f$ have been calculated including the effects of off-shell
gauge bosons~\cite{Bredenstein:2006rh,Bredenstein:2006ha}.  The calculation 
was improved by including final-state photon radiation off the fermions, 
resummed using a structure function approach, as well as higher-order 
corrections due to the Higgs self-coupling relevant for large Higgs 
masses~\cite{Ghinculov:1995bz,Frink:1996sv}, which increase the decay width 
for a 120 GeV Higgs by about 0.2\% at NLO and by another 0.003\% at NNLO.
The complete calculation has been implemented in the Monte Carlo generator
{\tt PROPHECY4F}~\cite{Bredenstein:2006rh,Bredenstein:2006ha}.  For Higgs 
masses below 400 GeV, the theoretical uncertainty is
below 1\%; for our study we estimate the theory uncertainty 
in $\Gamma_{WW}$ and $\Gamma_{ZZ}$ in the intermediate Higgs mass range 
to be at the 0.5\% level~\cite{Dittmaier-private}.

\subsection{$H \to gg$}
\label{sec:Hgg}

The decay $H \to gg$ appears first at one-loop mediated by a quark
triangle.  The top quark loop dominates the amplitude because it
contributes proportional to the large top Yukawa coupling $y_t = g
m_t/\sqrt{2} m_W \sim 1$.  The bottom quark loop is also included in
{\tt HDECAY}; it reduces the $H \to gg$ partial width by about 11\%
for $m_H = 120$~GeV compared to including the top loop alone.  We note
that {\tt HDECAY} v.3.103~\cite{HDECAY} (downloaded June 2006)
includes the charm quark loop in the calculation of the MSSM $h^0 \to
gg$ partial width, but not in the calculation of the SM $H \to gg$
partial width.  For consistency, we add the charm loop into the SM $H
\to gg$ calculation; it reduces $\Gamma_{gg}$ by about 2\% for $m_H =
120$ GeV.\footnote{This number is for {\tt NF-GG = 5}, i.e., the gluon
splitting to $b \bar b$ or $c \bar c$ at NLO is \emph{not} subtracted
off $\Gamma_{gg}$ and added on to $\Gamma_{b\bar b, c\bar c}$; see the
end of this subsection for further discussion of this issue.}

The QCD corrections to $H \to gg$ are most easily calculated in the
heavy quark limit, in which $m_t \to \infty$ is taken in 
the leading-order top triangle diagram (the amplitude goes to a constant
in this limit).  This allows the top triangle
to be shrunk to a point and replaced with an effective coupling of the
Higgs to two gluons.
{\tt HDECAY} includes the NLO QCD correction to $\Gamma_{gg}$
in this heavy quark limit only; in the intermediate Higgs mass
range this correction shifts $\Gamma_{gg}$
upward by about 60--65\% relative to the leading-order (LO) partial width.
The NLO QCD corrections with full dependence on the mass of the quark in
the loop are known~\cite{Spira:1995rr}; in the intermediate Higgs mass
range, including the full mass dependence shifts $\Gamma_{gg}$ upward
by an additional 5\% of the LO width~\cite{Spirareview}.  
This mass-dependent part of the correction is not included in {\tt HDECAY}.

The NNLO QCD correction is known only in the heavy quark
limit~\cite{Chetyrkin:1997iv}; it adds an additional 20\% to the LO result.
Very recently, the N$^3$LO QCD corrections to $\Gamma_{gg}$
have been computed~\cite{Baikov:2006ch}, again in the heavy quark limit; 
they add another 2\% to the LO result.
By varying the renormalization scale between $m_H/2$ and $2 m_H$,
a measure of the theoretical uncertainty of the calculation can be 
estimated: the scale dependence decreases from 24\% (LO) to 22\% (NLO)
to 10\% (NNLO), and finally to only 3\% at N$^3$LO~\cite{Baikov:2006ch}.
Neglecting the mass dependence of the NLO correction and
neglecting the 
NNLO and N$^3$LO corrections entirely leaves {\tt HDECAY} with about a 16\%
relative uncertainty on the calculated NLO partial width.

The electroweak corrections of $\mathcal{O}(G_F m_t^2)$ to $H \to gg$
were calculated in~\cite{Djouadi:1994ge} in the heavy-top approximation
and they give a simple multiplicative factor which increases the gluonic
decay width by about 0.3\%.  We thus expect the NNLO electroweak corrections
and the NLO finite-top-mass effects to be below the per-mille level.
This NLO electroweak correction is neglected in {\tt HDECAY}.

We thus take the remaining theory uncertainty in $\Gamma_{gg}$ to be roughly
3\%, corresponding to the scale uncertainty in the N$^3$LO QCD calculation.

Finally, we note here a potential issue for the theoretical interpretation
of Higgs partial width measurements that requires further study.
Theoretical calculations of the higher-order QCD corrections to $H \to gg$
include processes in which one of the final-state gluons splits into a 
quark-antiquark pair.  The {\tt HDECAY} code contains a switch that allows
the user to specify the number of final-state quark flavours included in 
$\Gamma_{gg}$; setting {\tt NF-GG = 5} includes the five light quark flavours
in $\Gamma_{gg}$, while {\tt NF-GG = 4} separates off the contribution from
gluon splitting to $b \bar b$ and adds it in to $\Gamma_{b \bar b}$.
{\tt NF-GG = 3} does the same for gluon splitting to $c \bar c$ as well.

Compared to {\tt NF-GG = 5} and for $m_H = 120$~GeV, 
subtracting off the gluon splitting to 
$b \bar b$ and adding it to $\Gamma_{b \bar b}$ (i.e., {\tt NF-GG = 4})
raises $\Gamma_{b \bar b}$ by about 1\% and reduces $\Gamma_{gg}$ by about
9\%.  Subtracting off the gluon splitting to $c \bar c$ and adding it to
$\Gamma_{c \bar c}$ (i.e., {\tt NF-GG = 3}) raises $\Gamma_{c \bar c}$ by
almost 30\% and reduces $\Gamma_{gg}$ by a further 12\%.
These effects are quite significant compared to the expected experimental
precision in the $gg$ and $c \bar c$ channels.
The amount of this gluon splitting correction that should be subtracted off
of $\Gamma_{gg}$ and included in $\Gamma_{b \bar b}$ or $\Gamma_{c \bar c}$
is an experimental question requiring a more detailed simulation of flavour
tagging of Higgs decays into three-body final states.  In our study we 
set {\tt NF-GG = 5} throughout and ignore this issue.

\subsection{$H \to \gamma\gamma$}

The SM $H \to \gamma\gamma$ decay partial width receives QCD corrections
to the one-loop diagrams involving quarks.  Because the 
external particles in the $H \gamma\gamma$ vertex are colour neutral, 
the virtual QCD corrections are finite by themselves.
The QCD corrections to the top quark loop in $\Gamma_{\gamma\gamma}$
are known analytically at NLO~\cite{ggH2loopQCD} 
and as a power expansion up to third order in $m_H/m_t$ at 
NNLO~\cite{ggH3loopQCD}.  The NLO corrections
are of order 2\% for $m_H < 2 m_W$, and the NNLO
corrections are at the level of a few per mille. 
These QCD corrections are neglected~\cite{Spirareview} in {\tt HDECAY}.

$\Gamma_{\gamma\gamma}$ also receives electroweak
corrections, the complete set of which have been calculated at NLO neglecting
the Yukawa couplings of all fermions except the top quark.
The complete calculation consists of the two-loop diagrams 
containing light fermion loops and $W$ or $Z$ bosons 
(with the Higgs coupled to the $W$ or $Z$, because the light 
fermion Yukawa couplings are neglected)~\cite{Aglietti};
the two-loop diagrams involving third-generation 
quarks and electroweak bosons~\cite{ggHEWGFmt2,ggHEWGFmt2old,Maltoni};
and the purely bosonic corrections~\cite{Maltoni}.
Together, these NLO electroweak corrections are between $-4\%$ and $0\%$
for 100 GeV $\lesssim m_H \lesssim 150$ GeV.
These electroweak corrections are neglected~\cite{Spirareview}
in {\tt HDECAY}.

The corrections to the part of the $H \to \gamma \gamma$ amplitude
involving the Higgs coupling to light quarks (e.g., $b$) have not been 
computed.  However, since the leading order contribution of the $b$ quark 
loop to $\Gamma_{\gamma\gamma}$ is at the few percent level, and 
the QCD and electroweak radiative corrections are themselves at the few 
percent level of the leading order piece, we estimate these uncalculated
two-loop corrections to be at most at the per-mille level.
We thus take the remaining theoretical uncertainty in 
$\Gamma_{\gamma\gamma}$ to be of order 0.1\%.

\subsection{$H \to Z \gamma$}

Like $H \to \gamma\gamma$,
the SM $H \to Z\gamma$ decay partial width receives QCD corrections to 
the one-loop diagrams involving quarks.
These corrections were
given in Ref.~\cite{Spira:1991tj} including the full dependence on the 
Higgs, $Z$ and top quark masses.  They amount to less than
0.3\% in the intermediate Higgs mass range~\cite{Spirareview} and have
been neglected in {\tt HDECAY}.

In analogy with the $H \to \gamma\gamma$ process, we expect the two-loop
electroweak corrections to $\Gamma_{Z\gamma}$ to be at the few-percent 
level; we thus take an overall theory uncertainty of 4\% for the
$H \to Z \gamma$ mode.

\section{Theoretical uncertainty in the $e^+e^- \to \nu \bar \nu H$ production 
cross section}
\label{app:xsec}

The full one-loop electroweak radiative corrections to Higgs production via 
$e^+e^- \to \nu \bar \nu H$ have been calculated in 
Refs.~\cite{Belanger,Dittmaier1,Dittmaier2}.  For the experimental selection
used in the 1000 GeV centre-of-mass energy ILC studies~\cite{Barklow}, 
the processes
$e^+e^- \to \nu_{\ell} \bar \nu_{\ell} H$ ($\ell = e, \mu, \tau$) arising
from both $WW$ fusion and Higgsstrahlung contribute, with $WW$ fusion
dominating.
The one-loop electroweak corrections to these processes
reduce the cross section by about 
10\%~\cite{Dittmaier1,Dittmaier2} in the parameter range of interest to us. 

Beyond one loop, the $e^+e^- \to \nu \bar \nu H$ cross section will receive
additional electroweak and QCD corrections.  
We estimate the missing two-loop electroweak 
corrections to be roughly of order (one-loop)$^2$.
The largest electroweak corrections are due to initial-state radiation, 
which is already computed in the leading-log approximation up to order 
$\alpha^3$~\cite{Dittmaier2}.  The remaining purely weak one-loop corrections
are of order 3--4\%~\cite{Dittmaier2}, leading us to estimate a 
(one-loop)$^2$ uncertainty of about 0.1--0.2\%.  
At two loops, QCD corrections
will affect only the fermionic component of the one-loop electroweak 
corrections; this fermionic 
component amounts to about a 2\% reduction of the total
cross section in our parameter region of interest~\cite{Dittmaier1}, so
we estimate the QCD correction at two loops to be of order 0.2\%.  
At TeV energies one should also consider the enhanced effects that 
often occur in high-energy ($E \gg m_W$) processes, such as logarithmic
running of couplings and soft or collinear weak boson exchange; these
effects can be at the percent level even at two loops.  
However, the $t$-channel weak boson 
fusion process of interest here is dominated by small energy transfers, so 
that we do not expect these effects to exceed the 0.1\% level.
For a relatively light Higgs mass $m_H \sim 120$--140 GeV the Higgs 
self-coupling is relatively small, so that we should also be safe from 
large corrections due to Higgs self-interaction diagrams at the two-loop
level.

We thus estimate a combined theory uncertainty on the $e^+e^- \to \nu
\bar \nu H$ cross section at 1000 GeV centre-of-mass energy of
0.5\%~\cite{Dittmaier-private}.  We emphasize that this is just a
``reasonable'' estimate; to make this number really solid would
require a serious study of the size of the two-loop effects mentioned
above.

\section{Uncertainties in $\overline{m_b}(\overline{m_b})$ and 
$\overline{m_c}(\overline{m_c})$}
\label{app:mbmc}

The predictions for the Higgs partial widths into bottom or charm quark 
pairs are computed up to three loops in terms of the $\overline{\rm MS}$
quark masses evaluated at the scale of the Higgs mass.  As input, the 
calculation requires the bottom and charm quark $\overline{\rm MS}$ masses
at a starting scale such as the quark's own mass: 
$\overline{m_b}(\overline{m_b})$ and $\overline{m_c}(\overline{m_c})$.
We summarize here the various ways that $\overline{m_b}(\overline{m_b})$ and 
$\overline{m_c}(\overline{m_c})$ have been extracted; the masses obtained are 
summarized in Tables~\ref{tab:mc} and \ref{tab:mb}.

\begin{table}
\begin{tabular}{ll}
\hline\hline
Source & $\overline{m_c}(\overline{m_c})$ [GeV] \\
\hline
Inclusive semileptonic $B$ decays & \\
\hspace*{1cm} via $m_b^{\rm 1S}$ and $m_b-m_c$ \cite{HoangBsl} & 
     $1.224 \pm 0.057$ 
     ($*$) \\
\hspace*{1cm} via kinetic masses \cite{Buchmuller} &
     $1.24 \pm 0.07$ \\
$e^+e^- \to$ hadrons & \\
\hspace*{1cm} from spectral moments \cite{Kuhneehad} & 
     $1.304 \pm 0.027$ \\
\hspace*{1cm} more conservative error analysis \cite{Hoangeehad} &
     $1.29 \pm 0.07$ \\
Quenched lattice QCD & \\
\hspace*{1cm} from $F_K$ and $D_s$ mass \cite{RolfLattice} &
     $1.301 \pm 0.034$ $\pm$ ?? \\
\hspace*{1cm} from $\bar c c$ and $\bar c s$ state masses 
     \cite{DivitiisLattice} &
     $1.319 \pm 0.028$ $\pm$ ?? \\
Unquenched lattice QCD & \\
\hspace*{1cm} meson spectra (prelim) \cite{NobesLattice} &
     $1.22 \pm 0.09$ \\
\hline\hline
\end{tabular}
\caption{$\overline{\rm MS}$ charm quark masses, 
$\overline{m_c}(\overline{m_c})$, from recent analyses (see text for details).
In our study we use the charm mass extracted from inclusive 
semileptonic $B$ decays~\cite{HoangBsl}, marked with an asterisk ($*$) 
in the table.  For quenched lattice QCD the question marks ?? denote the 
uncontrolled uncertainty due to the quenched approximation.}
\label{tab:mc}
\end{table}

\begin{table}
\begin{tabular}{ll}
\hline\hline
Source & $\overline{m_b}(\overline{m_b})$ [GeV] \\
\hline
Inclusive semileptonic $B$ decays & \\
\hspace*{1cm} via kinetic masses \cite{Buchmuller} &
     $4.20 \pm 0.04$ ($*$) \\
$e^+e^- \to$ hadrons & \\
\hspace*{1cm} from spectral moments \cite{Kuhneehad} & 
     $4.191 \pm 0.051$ \\
\hspace*{1cm} more conservative error analysis \cite{Hoangeehad} &
     $4.22 \pm 0.11$ \\
Quenched lattice QCD & \\
\hspace*{1cm} from $\bar b b$ and $\bar b s$ state masses 
     \cite{DivitiisLattice} &
     $4.33 \pm 0.10$ $\pm$ ?? \\
Unquenched lattice QCD & \\
\hspace*{1cm} $\Upsilon$ spectrum \cite{GrayLattice} &
     $4.4 \pm 0.3$ \\
\hspace*{1cm} meson spectra (prelim) \cite{NobesLattice} &
     $4.7 \pm 0.4$ \\
\hline\hline
\end{tabular}
\caption{$\overline{\rm MS}$ bottom quark masses, 
$\overline{m_b}(\overline{m_b})$, from recent analyses (see text for details).
In our study we use the bottom mass extracted from inclusive 
semileptonic $B$ decays~\cite{Buchmuller}, marked with an asterisk ($*$) 
in the table.  For quenched lattice QCD the question marks ?? denote the 
uncontrolled uncertainty due to the quenched approximation.}
\label{tab:mb}
\end{table}

Extracting the precise values of the charm and bottom quark masses from
experiment is a challenge because of QCD effects which are starting to
become strong at the bottom and charm mass scales.  In particular, 
perturbative QCD does not allow one to define the quark pole masses with
an accuracy better than $\Lambda_{\rm QCD} \sim 200$ MeV; this is known
as the renormalon ambiguity.\footnote{A renormalon is a power-like 
infrared-sensitive contribution to a hard process -- a relative 
correction of order $\Lambda_{\rm QCD}/m_Q$ in our case -- as opposed 
to the usual logarithmic infrared-sensitive effects 
$\log(\Lambda_{\rm QCD}/m_Q)$ encountered in QCD calculations.}  Because
the $\overline{\rm MS}$ quark masses are short-distance mass definitions, 
they are in principle free of this ambiguity.  

\subsection{Inclusive semileptonic $B$ decays}

The charm mass has recently been extracted~\cite{HoangBsl} from a global fit 
to $B$ meson decay spectra~\cite{BauerBsl}.  The procedure combines the
values of $m_b-m_c$ and $m_b^{\rm 1S}$ from a fit of the decay spectra
of $B \to X_c \ell \bar \nu$ and $B \to X_s \gamma$ to a Heavy Quark 
Effective Theory (HQET) expansion performed in Ref.~\cite{BauerBsl}.  The 
proper treatment of the renormalon cancellation was done in 
Ref.~\cite{HoangBsl} to extract the charm mass 
$\overline{m_c}(\overline{m_c}) = 1.224 \pm 0.017_{\rm expt} 
\pm 0.054_{\rm thy}$ GeV, where the first error is experimental and
includes the uncertainty in $\alpha_s$ and the second error represents
a conservative combination of the theory uncertainties in the 
extraction~\cite{HoangBsl}.  While this theory uncertainty is not meant
to be interpreted as a statistical uncertainty (it represents a range where
the true charm mass should be located with probability much higher than
the 67\% probability of a 1$\sigma$ interval), we will nevertheless treat
it as a 1$\sigma$ Gaussian error and combine it in quadrature with the
experimental error for inclusion in our $\chi^2$ fits.  For our analysis
we thus take a charm mass of 
\begin{equation}
  \overline{m_c}(\overline{m_c}) = 1.224 \pm 0.057 \ {\rm GeV}.
\end{equation}
The quoted uncertainty amounts to a 4.7\% relative uncertainty, which
is dominated by the theoretical uncertainty.  The 
theoretical uncertainty could be improved in the future by a full 
$\mathcal{O}(\alpha_s^2)$ analysis of the inclusive $B$ decay spectra 
used in the extraction, to reduce the theoretical uncertainty in the fit
to the HQET parameters.

The charm quark mass has also been extracted in Ref.~\cite{Buchmuller} 
through a similar fit to the $B \to X_c \ell \bar \nu$ and $B \to X_s \gamma$
decay distributions using HQET expansions in the kinetic scheme, yielding 
kinetic masses $m_b$ and $m_c$ which are converted to $\overline{\rm MS}$ 
masses using a formula from Ref.~\cite{Bigi}, accurate to two loops 
and including part of the three-loop $\alpha_s^3$ corrections.  The resulting
$\overline{\rm MS}$ charm mass is
$\overline{m_c}(\overline{m_c}) = 1.24 \pm 0.07$ GeV.  This is consistent
within the uncertainties with the charm mass found in Ref.~\cite{HoangBsl}.

The fits of the $B \to X_c \ell \bar \nu$ and $B \to X_s \gamma$
decay distributions to HQET expansions are also used to extract the bottom
quark mass.  The extraction can be done using any of a variety of different 
schemes for the HQET expansions~\cite{BauerBsl}; however, some schemes are 
better than others in the sense that they suffer from smaller theoretical 
uncertainties.  Currently the smallest uncertainty comes with the bottom
mass extraction in the $1S$ scheme, 
$m_b^{1S} = 4.68 \pm 0.03$ GeV~\cite{BauerBsl}, 
an uncertainty of only 0.64\%.
The bottom mass has also been extracted~\cite{BauerBsl,Buchmuller} 
in the kinetic scheme discussed above for the charm mass extraction.
As for the charm quark, the kinetic bottom quark mass can be translated 
into an $\overline{\rm MS}$ mass using the formula from Ref.~\cite{Bigi}.
Following this procedure, 
Ref.~\cite{Buchmuller} finds for the $\overline{\rm MS}$ mass,
\begin{equation}
  \overline{m_b}(\overline{m_b}) = 4.20 \pm 0.04 \ {\rm GeV}.
\end{equation}
We use this mass in our analysis.  The relative uncertainty is 0.95\%, and is 
dominated by the theoretical uncertainties that feed into the fit errors.

\subsection{$e^+e^- \to$ hadrons}

The inclusive cross section for $e^+e^- \to$ hadrons as a function of the
centre-of-mass energy is sensitive to the fundamental parameters of QCD, 
including the quark masses.  In particular, precise perturbative QCD 
calculations of the first few ``moments'' of the cross section, which depend
heavily on the threshold region, allow a precise determination of the 
$\overline{\rm MS}$ charm and bottom quark masses.

The moments $P_n = \int R_{q \bar q}(s) s^{-(n+1)} ds$, where 
$R_{q \bar q} = \sigma(e^+e^- \to q \bar q + X) / 
\sigma(e^+e^- \to \mu^+\mu^-)$ and $q = c$ or $b$, are obtained 
experimentally from measurements of the production cross section 
near the $c \bar c$ and $b \bar b$ thresholds.  Experimental uncertainties
include gaps in the data above the quarkonium resonance regions and 
uncertainties in the large contributions from the narrow quarkonium 
resonances~\cite{Corcellaeehad}.
The quark masses are extracted by matching the experimentally determined
moments to theoretical predictions.  The uncertainties on the theory side
include the unknown higher-order QCD contributions to the moments, which
manifest both as scale dependence within a single scheme and differing 
results for the quark masses depending on how the renormalization scale
is chosen as a function of the $e^+e^-$ centre-of-mass 
energy~\cite{Hoangeehad}.

A recent fit~\cite{Kuhneehad} found 
$\overline{m_c}(\overline{m_c}) = 1.304 \pm 0.027$ GeV and 
$\overline{m_b}(\overline{m_b}) = 4.191 \pm 0.051$ GeV.  This remarkably 
small error estimate for $m_c$ was reconsidered in 
Refs.~\cite{Corcellaeehad,Hoangeehad}, with special attention being paid
to sources of theory uncertainty, resulting in revised mass extractions of 
$\overline{m_c}(\overline{m_c}) = 1.29 \pm 0.07$ GeV and 
$\overline{m_b}(\overline{m_b}) = 4.22 \pm 0.11$ GeV~\cite{Hoangeehad}.

Very recently the first moment of the hadronic production
cross section was computed at 
$\mathcal{O}(\alpha_s^3)$~\cite{Chetyrkineehad,Boughezaleehad}, significantly
reducing the renormalization scale dependence.  Repeating
the analysis of Ref.~\cite{Kuhneehad}, Ref.~\cite{Boughezaleehad} finds
$\overline{m_c}(\overline{m_c}) = 1.295 \pm 0.015$ GeV and 
$\overline{m_b}(\overline{m_b}) = 4.205 \pm 0.058$ GeV.  We await a more
detailed examination of the theoretical uncertainties a la 
Refs.~\cite{Corcellaeehad,Hoangeehad} before making use of this very small
uncertainty in $m_c$.

\subsection{Lattice QCD}

Lattice QCD allows the extraction of the quark masses through fits of
the fundamental QCD parameters -- $\alpha_s$ and the quark masses --
to precisely-measured meson properties such as the kaon mass and decay
constant $F_K$ (to set the QCD scale and the light quark masses) and
the mass of the $D_s$ meson (to provide sensitivity to the charm quark
mass)~\cite{RolfLattice}.  Adding $B$ mesons to the simulation allows
access to the bottom quark mass \cite{DivitiisLattice}.  The charm and
bottom masses have been thus calculated in \emph{quenched} lattice
QCD, in which only virtual gluons are simulated and not virtual
light-quark--antiquark pairs.  Recent analyses find
$\overline{m_c}(\overline{m_c}) = 1.301 \pm 0.034$ GeV
\cite{RolfLattice} or $1.319 \pm 0.028$ GeV \cite{DivitiisLattice} and
$\overline{m_b}(\overline{m_b}) = 4.33 \pm 0.10$ GeV
\cite{DivitiisLattice}.  However, quenched lattice QCD calculations
suffer from an uncontrolled theoretical uncertainty due to the
quenched approximation.

Unquenched lattice QCD simulations are significantly more
computer-intensive and have only recently yielded significant results.
The bottom quark mass in full (unquenched) lattice QCD was obtained
for the first time in Ref.~\cite{GrayLattice} from the $\Upsilon$
spectrum.  The bare mass on the lattice was extracted with 1--2\%
uncertainty, which is mostly systematic and due to missing radiative
corrections to the leading relativistic and discretization corrections
to the $\Upsilon$ binding energies.  The bare mass on the lattice
depends on the lattice spacing, which cuts off the effective theory on
the lattice; it can be converted to the $\overline{\rm MS}$ mass
without encountering infrared problems.  The conversion is done at one
loop yielding $\overline{m_b}(\overline{m_b}) = 4.4 \pm 0.3$
GeV~\cite{GrayLattice}, where the error is dominated by the
perturbative uncertainty in the conversion from the bare lattice mass.
A two-loop calculation of the heavy quark self-energy to reduce the
conversion error is underway~\cite{GrayLattice}.  Reducing the error
in $\overline{m_b}(\overline{m_b})$ to the level of the uncertainty in
the bare lattice mass would make the lattice determination of
$\overline{m_b}(\overline{m_b})$ competitive with that from inclusive
$B$ meson decay distributions discussed above.  A similar
(preliminary) analysis~\cite{NobesLattice} including also the charm
quark gives $\overline{m_c}(\overline{m_c}) = 1.22 \pm 0.09$ GeV and
$\overline{m_b}(\overline{m_b}) = 4.7 \pm 0.4$ GeV.  These masses are
all in good agreement within their uncertainties with the charm and
bottom masses found through other techniques.


\end{document}